\pdfoutput=1

\documentclass[11pt]{article}

\usepackage[preprint]{acl}
\usepackage{enumerate}

\usepackage{booktabs}  
\usepackage{pifont}    
\usepackage{multirow}  
\usepackage{adjustbox} 
\usepackage{times}
\usepackage{latexsym}
\usepackage{booktabs}    
\usepackage{graphicx}    
\usepackage{fontawesome} 
\usepackage{array}       
\usepackage{multirow}    
\usepackage{colortbl}    
\usepackage{ulem} 
\usepackage[T1]{fontenc}

\usepackage[utf8]{inputenc}

\usepackage{microtype}

\usepackage{inconsolata}
\usepackage{multirow}
\usepackage{graphicx}

\usepackage{bbm}
\usepackage{colortbl}
\definecolor{mypink}{rgb}{.99,.91,.95}
\definecolor{mygreen}{rgb}{.9,.99,.9}
\definecolor{mygray}{gray}{.9}
\usepackage{xspace}
\usepackage{wrapfig}
\usepackage{arydshln}
\usepackage{subcaption} 
\usepackage{makecell}
\usepackage{tcolorbox}
\tcbuselibrary{breakable}
\usepackage{listings}

\usepackage{hyperref}       
\usepackage{url}            
\usepackage{booktabs}       
\usepackage{amsfonts}       
\usepackage{nicefrac}       
\usepackage{microtype}      
\usepackage{xcolor}         
\usepackage{xspace}
\usepackage{wrapfig}
\usepackage{amsmath}
\usepackage{arydshln}
\usepackage{makecell}
\usepackage{tcolorbox}
\usepackage{listings}

\usepackage{graphicx}
\usepackage{pifont} 
\usepackage{CJKutf8}
\usepackage{paralist}
\usepackage{subcaption}

\newcommand{\resultone}[1]{\colorbox{green!15}{#1}}
\newcommand{\resulttwo}[1]{\colorbox{cyan!15}{#1}}
\newcommand{\resultthird}[1]{\colorbox{yellow!15}{#1}}

\newcommand{\ourmethod}{\textsc{\textbf{DependEval}}}

\newtcolorbox{mybox}[1][]{
  arc=1mm,
  boxrule=1pt,
  colback=gray!20, 
  colframe=black!80,
  fonttitle=\bfseries,
  fontupper=\small, 
  title=#1,
  left=1mm,
  right=1mm,
  top=1mm,
  bottom=1mm
}
\usepackage{tikz}

\title{
    \makebox[0pt][l]{\hspace{-2.4em}\raisebox{-0.5em}{\includegraphics[width=0.07\textwidth]{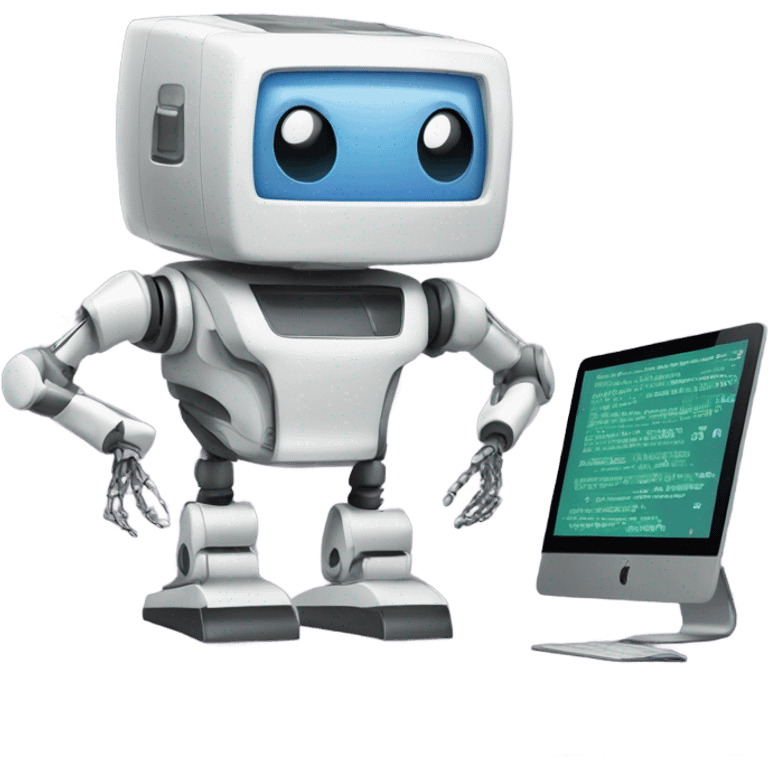}}}
    DependEval: Benchmarking LLMs for Repository Dependency Understanding
}

\author{
 \textbf{Junjia Du\textsuperscript{1}},
 \textbf{Yadi Liu\textsuperscript{2}},
 \textbf{Hongcheng Guo\textsuperscript{3}}\thanks{Corresponding author.}, 
 \textbf{Jiawei Wang\textsuperscript{4}},
\\
 \textbf{Haojian Huang\textsuperscript{5}},
 \textbf{Yunyi Ni\textsuperscript{1}},
 \textbf{Zhoujun Li\textsuperscript{3}},
\\
\\
 \textsuperscript{1}Nanyang Technological University 
 \textsuperscript{2}Tsinghua University,
 \textsuperscript{3}Beihang University \\
 \textsuperscript{4}ShanghaiTech University 
 \textsuperscript{5}The University of Hong Kong
\\
}
\lstdefinelanguage{json}{
    basicstyle=\ttfamily\small,
    commentstyle=\color{gray},
    stringstyle=\color{black},
    keywordstyle=\color{blue},
    morestring=[b]",
    morestring=[s]{:}{,},
    morecomment=[l]{//},
    morecomment=[s]{/*}{*/},
    morekeywords={true,false,null}
}

\begin{document}
\maketitle
\begin{abstract}

While large language models (LLMs) have shown considerable promise in code generation, real-world software development demands advanced repository-level reasoning. This includes understanding dependencies, project structures, and managing multi-file changes. However, the ability of LLMs to effectively comprehend and handle complex code repositories has yet to be fully explored. To address challenges, we introduce a hierarchical benchmark designed to evaluate repository \textbf{dependency} understanding (\textbf{\ourmethod{}}). Benchmark is based on 15,576 repositories collected from real-world websites. It evaluates models on three core tasks: \textbf{Dependency Recognition}, \textbf{Repository Construction}, and \textbf{Multi-file Editing}, across 8 programming languages from actual code repositories. Our evaluation of over 25 LLMs reveals substantial performance gaps and provides valuable insights into repository-level code understanding\footnote{We provide code and datasets at \href{https://github.com/ink7-sudo/DependEval}{https://github.com/ink7-sudo/DependEval}.}

\end{abstract}










\section{Introduction}

\begin{quote} \large\textit{``It’s not that we use a language to express our thoughts, but that we use a language to understand our thoughts.''} \textbf{-- Donald Knuth} \end{quote}

Large Language Models (LLMs) have made significant strides in automated code generation and comprehension~\cite{code_bert,code_llama,codealpaca,qwen25coder,codegen2}, enabling tasks like code completion, bug fixing, and documentation. However, real-world software development involves complex repository-wide reasoning, including dependency understanding, project structures, and multi-file modifications~\cite{repoformer,zhang2023repocoder,shrivastava2023repofusion}. While LLMs show potential in software engineering, their ability to handle large-scale repositories remains underexplored.

Existing benchmarks~\cite{humaneval_xl,mceval,evalplus,multipl_e} typically evaluate LLMs at the function or file level, often focusing on isolated code snippets. However, large software projects require an evaluation framework that captures repository-wide reasoning. Previous benchmarks, such as SWE-bench~\cite{swe_bench}, focus on bug fixing but neglect challenges like dependency resolution, project construction, and structured multi-file edits. Additionally, existing methods~\cite{repobench,repoeval,ding2023cceval} lack fine-grained insights into model performance across repository reasoning aspects.

To address these limitations, we introduce \ourmethod{}, a hierarchical benchmark designed to evaluate LLMs’ ability to reason about structured code repositories through three progressively challenging tasks:  
(1) \textbf{Dependency Recognition}: Assessing the model’s ability to infer and resolve inter-file dependencies.  
(2) \textbf{Repository Construction}: Evaluating models on generating structured, modular project layouts.  
(3) \textbf{Multi-File Editing}: Testing models' ability to make coordinated changes across multiple files while maintaining consistency. Unlike existing benchmarks~\cite{ding2023cceval,repobench}, \ourmethod{} provides fine-grained metrics for static analysis, architectural reasoning, and cross-file consistency.

We evaluate \ourmethod{} with 25+ LLMs, making the following key contributions:  

\begin{itemize}
    \item \ourmethod{} is the first multilingual benchmark with over 2,500 cases designed for hierarchical assessment of repository-scale understanding in LLMs, enabling a comprehensive evaluation of LLMs' understanding capabilities at the code repository level.
    \item It provides fine-grained metrics, enabling detailed analysis of repository reasoning capabilities.  
    \item Larger models tend to perform better, particularly on complex cross-file reasoning tasks, but performance gains diminish as task complexity increases, suggesting that scaling alone is not enough for mastering repository-wide code comprehension.  
    \item Experiment analysis shows that LLMs struggle with key challenges in large-scale software development, such as dependency parsing, function call inference, and maintaining consistency across file modifications, highlighting key challenges in applying LLMs to large-scale software development.  
\end{itemize}


\section{Related Work}

\noindent\textbf{Code Large Language Models.}
The rapid advancement of generative language models has spurred extensive research on AI applications in software engineering~\citep{black2022gptneox, brown2020gpt3, radford2019gpt2, touvron2023llama}. Models~\citep{achiam2023gpt4, chen2021evaluating, chowdhery2023palm,nijkamp2023codegen2, nijkamp2022codegen,li2023starcoder, lozhkov2024starcoder2,roziere2023codellama,guo2024deepseek} have demonstrated significant performance improvements on various code-related tasks. These advancements have accelerated the development of code assistant tools like Copilot~\footnote{https://github.com/features/copilot}, TONGYI Lingma~\cite{qwen25coder}, and Cursor~\footnote{https://www.cursor.com/}.

\noindent\textbf{Repository-level Code Evaluation.}
Recent repository-level code benchmarks~\citep{allal2023santacoder, liu2023repobench, shrivastava2022repository, zhang2023repocoder, ding2023cceval, Liu2024M2rcEvalMM, allal2023santacoder, repobench} assess code LLMs' capabilities at various granularities, from individual lines and API calls to full function implementations. SWE-Bench~\cite{swe_bench} challenges LLMs with real-world scenarios using issues and pull requests from popular Python repositories on GitHub, while DevBench~\cite{DevBench} breaks the development process into stages to evaluate AI performance at each step. However, these datasets often neglect detailed repository code understanding. Most difficulty categorizations focus on the number of files involved, overlooking the internal structure and semantic context within projects. To better evaluate multilingual, repository-based code understanding, \ourmethod{} extends to 8 programming languages and includes three designed tasks. The detailed comparison is in Table~\ref{table:comparison}.

\begin{table}[htbp]
\centering
\resizebox{\linewidth}{!}{%
\begin{tabular}{lllcl}
\toprule
\textbf{Benchmark} & \textbf{Task}                              & \textbf{Scope}      & \textbf{Languages} & \textbf{\#Repo} \\ \midrule
MBPP~\cite{mbpp}                     & Code Generation         & Function   & 1 & N/A   \\
HumanEval~\cite{evalplus}           & Code Generation         & Function   & 1 & N/A   \\
ClassEval~\cite{classeval}           & Code Generation         & Class      & 1 & N/A   \\
RepoEval~\cite{repocoder}            & Code Completion         & Repo-level & 1 & 14    \\
RepoBench~\cite{repobench}           & Retrieval \& Completion & Repo-level & 2 & 1,669 \\
CrossCodeEval~\cite{ding2023cceval}          & Code Completion         & Repo-level & 4 & 1,002 \\
EvoCodeBench~\cite{EvoCodeBench}     & Code Generation         & Repo-level & 1 & 25    \\
RepoMasterEval~\cite{RepoMasterEval} & Code Completion         & Repo-level & 2 & 6     \\ \midrule
\rowcolor{gray!15} \ourmethod{}         & \textbf{Hierarchical Dependency Reasoning} & \textbf{Repo-level} & \textbf{8}         & \textbf{2,683}  \\ \bottomrule
\end{tabular}%
}
\caption{Comparison of features between existing benchmarks and \ourmethod{}}
\label{table:comparison}
\end{table}

\section{DependEval}

\begin{figure*}[htbp]
    \centering
    \adjustbox{center}{\includegraphics[width=1.1\textwidth]{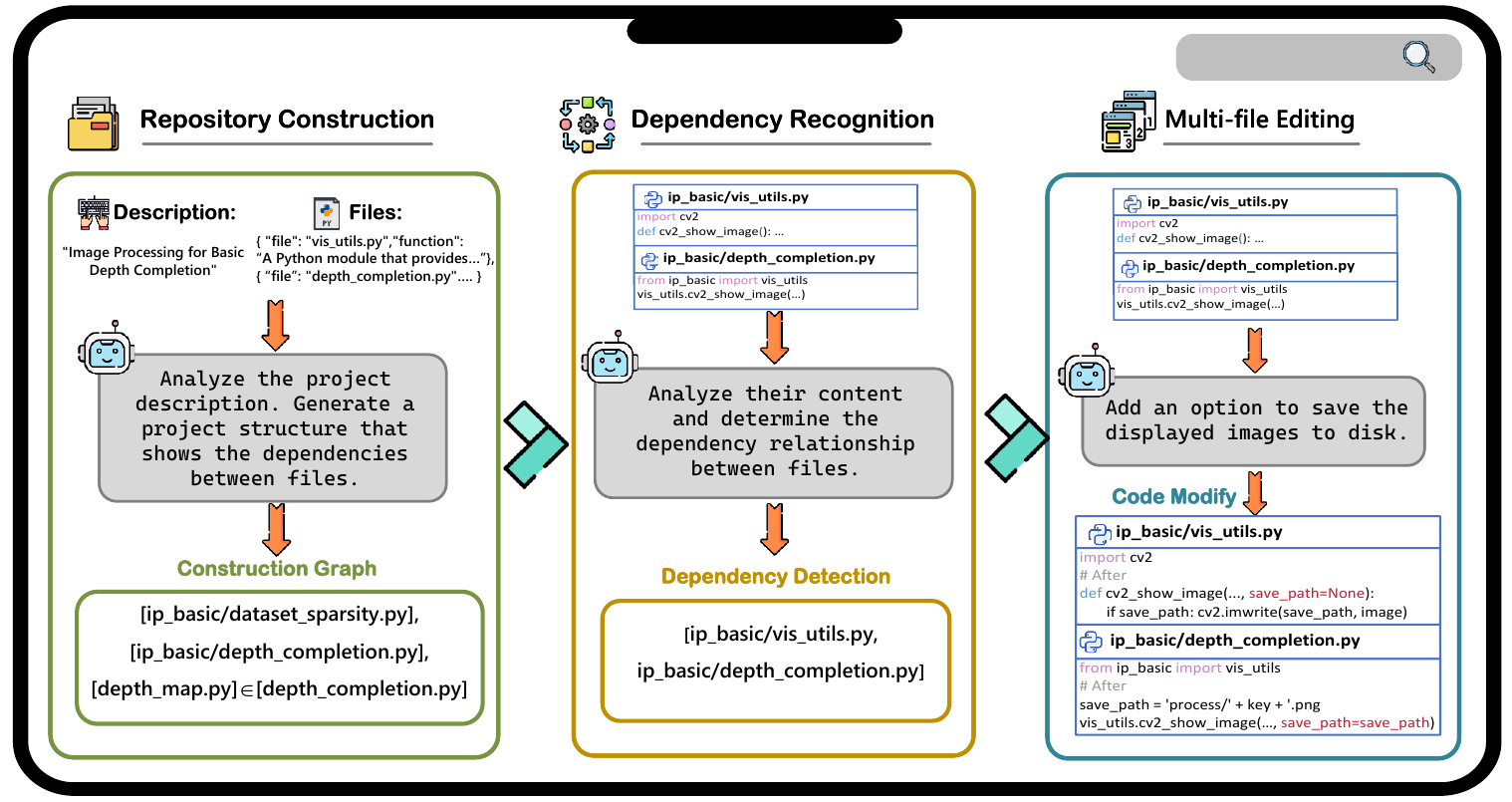}}
    \caption{Overview of \ourmethod{}. It contains 3 tasks including Repository Construction, Dependency Recognition, and Multi-file Editing. The first task analyzes the project description to generate a structure showing the dependencies between files. The second task identifies the content of the files to determine the relationships between them. Finally, the third task modifies the code to add functionality for saving displayed images to disk.}
    \label{fig:taskcase}
\end{figure*}

\subsection{Overview}
DependEval is a multi-language benchmark spanning eight programming languages (C, C++, C\#, TypeScript, JavaScript, Java, PHP, and Python) to evaluate the capability of large language models (LLMs) in three  hierarchical understanding tasks in Figure~\ref{fig:taskcase}.

\subsection{Data Collection and Filtering}\label{filtering}
Our data comes from real-world GitHub repositories, ensuring authenticity and practical relevance for evaluating code understanding tasks. We collect multilingual repositories in languages such as C, C++, C\#, TypeScript, JavaScript, Java, PHP, and Python, using GitHub's official API to crawl repositories and their README files. These are selected from open-source repositories created before December 16, 2024, with a focus on those with the highest star counts (e.g., above 50 stars) and excluding forks to mitigate data leakage and ensure quality. We apply filtering rules similar to Deepseek-Coder~\citep{deepseek_coder} to remove lower-quality code. Additionally, we filter READMEs based on structural integrity, practical utility, and content clarity. Detailed filtering criteria can be seen in Appendix~\ref{appendix:filter}. After filtering and removing duplicates, we retain 15,576 high-quality repositories and README files.

\subsection{Dependency Code Snippets Generation}\label{sec:snippets}

After downloading repositories from GitHub, we concatenate interdependent code files into snippets based on their dependency relationships. This involves parsing each file for import statements to identify internal module dependencies. Details on the import expressions considered for each language are provided in Appendix~\ref{appendix:Import statement}. Call chains and code snippets are generated by selecting a starting file and appending its dependent files in reference order. Multiple call chains of varying lengths can be extracted from each repository, capturing both short and long inter-file relationships. In addition to the filtering criteria in Section~\ref{filtering}, we exclude repositories that fail to cover all dependencies or do not pass integrity checks, ensuring only valid, self-contained code fragments are included.

\begin{figure*}[htbp]
    \centering
    \adjustbox{center}{\includegraphics[width=1.1\textwidth]{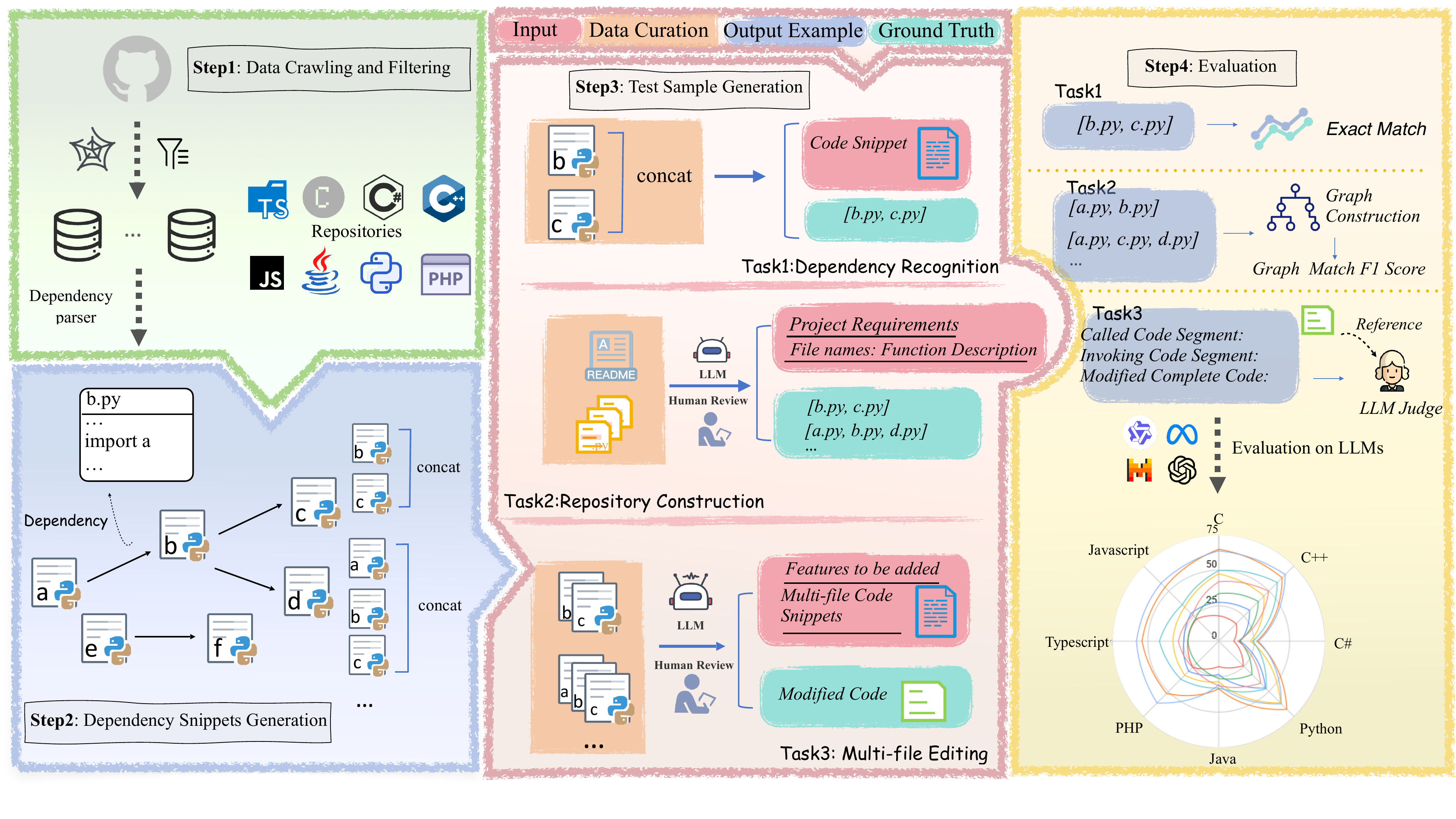}}
    \caption{Pipeline for data curation of \ourmethod{}. It consists of four steps: data crawling and filtering (Step 1), dependency snippet generation (Step 2), test sample generation for dependency recognition (Step 3), and evaluation using metrics like Exact Match and Graph Match F1 Score (Step 4).}
    \label{fig:datacuration}
\end{figure*}

\subsection{Task Construction}   

\subsubsection{Dependency Recognition}
\paragraph{Task Settings.}

The Dependency Recognition task is designed to evaluate the model's ability to accurately identify and understand the calling relationships between files within a codebase. Given a set of code files \( \mathcal{F} \), with dependency relations \( \mathcal{D} \) extracted as described in Section~\ref{sec:snippets}, we define the set of dependencies as:
\begin{equation}
\mathcal{D} \stackrel{\triangle}{=} \{(f_i, f_j) \mid f_i \text{ directly invokes } f_j\}.
\end{equation}

The model is required to generate a unique ordered list \( P = [p_1, p_2, \dots, p_n] \), where \( p_i \in \mathcal{F} \), such that:
\begin{equation}
\forall (f_i, f_j) \in \mathcal{D}, \ \pi^{-1}(f_j) < \pi^{-1}(f_i),
\end{equation}
where the output list \( P \) must respect all dependencies in \( \mathcal{D} \). Here, \( \pi^{-1}(f) \) denotes the position of file \( f \) in the list \( P \).

 
\subsubsection{Repository Construction}

\paragraph{Task Settings.}
The Repository Construction task evaluates the ability of LLMs to generate a coherent file structure based on natural language requirements and file names with functional descriptions. Given natural language requirements \( R \) and a set of files \( \mathcal{F} = \{f_1, f_2, \ldots, f_n\} \) with functional descriptions \( D = \{d_1, d_2, \ldots, d_n\} \), the input $\mathcal{X}$ is:
\begin{equation}
\mathcal{X} \stackrel{\triangle}{=} (R, \{(f_i, d_i)\}_{i=1}^{n}),
\end{equation}
where \( R \) is the requirement, and each \( (f_i, d_i) \) pair represents a file \( f_i \) and its description \( d_i \). The model generates a set of valid ordered lists \( P = \{ P_1, P_2, \ldots, P_m \} \), where each list \( P_i = [p_{i1}, p_{i2}, \ldots, p_{in}] \) consists of files \( p_{ij} \in \mathcal{F} \), with each file \( p_{ij} \) being invoked or depended upon by the subsequent file \( p_{i(j+1)} \).
\paragraph{Sample Curation.}
The construction process begins by extracting complete dependency relationships from the repository to identify invocation chains, which serve as the ground truth for evaluation. We filter repositories to include only those with fewer than 12 invocation chains, ensuring all files are comprehensively covered. Next, GPT-4 processes the README and individual code files to generate a natural language requirement, including an overall repository description and functional descriptions for each file. Detailed prompt is in Appendix~\ref{prompt:datacuration}.
\paragraph{Human in the Loop.} 
Finally, human review selects the final set of samples. We evaluate the generated natural language requirements based on criteria such as the coherence and completeness of repository descriptions, accuracy of functional descriptions, and correctness of reflected dependency relationships. Detailed criteria are provided in Appendix~\ref{appendix:task4annotator}. We also ensure a diverse range of topic types in the selected repositories to represent real-world scenarios, with further details in Appendix~\ref{appendix:diversity}.

\begin{table*}[h]
\centering
\resizebox{\textwidth}{!}{%
\begin{tabular}{cccccccccccc}
\toprule
Task &
  \multicolumn{1}{l}{} &
  \multicolumn{8}{c}{\# Examples} &
  \multicolumn{1}{l}{Avg. \# Input Tokens} &
  \multicolumn{1}{l}{Avg. \# Output Tokens} \\ 
\textbf{} &
  \textbf{Subtask} &
  \textbf{C} &
  \textbf{C++} &
  \textbf{C\#} &
  \textbf{Java} &
  \textbf{PHP} &
  \textbf{Python} &
  \textbf{JavaScript} &
  \textbf{TypeScript} &
  \multicolumn{1}{l}{} &
  \multicolumn{1}{l}{} \\ \midrule
\multicolumn{1}{r}{\multirow{3}{*}{Multi-file Editing (In-place Edits)}} &
  Chain Length 2 &
  30 &
  38 &
  49 &
  42 &
  49 &
  60 &
  58 &
  42 &
  21453 &
  4832 \\
\multicolumn{1}{r}{} &
  Chain Length 3 &
  53 &
  35 &
  34 &
  33 &
  37 &
  50 &
  33 &
  33 &
  28634 &
  5424 \\
\multicolumn{1}{r}{} &
  Chain Length 4 &
  39 &
  38 &
  33 &
  33 &
  60 &
  24 &
  32 &
  28 &
  32423 &
  6367 \\ \midrule
\multirow{2}{*}{Multi-file Editing (Expansion Edits)} &
  Chain Length 2 &
  44 &
  45 &
  23 &
  42 &
  38 &
  45 &
  49 &
  42 &
  21355 &
  7319 \\
 &
  Chain Length 3 &
  35 &
  21 &
  54 &
  41 &
  33 &
  38 &
  36 &
  34 &
  25542 &
  7844 \\ \midrule
 &
  \textbf{Sub-total} &
  201 &
  177 &
  193 &
  191 &
  217 &
  217 &
  208 &
  187 &
  - &
  - \\
Dependency Recognition &
  - &
  180 &
  180 &
  180 &
  180 &
  180 &
  180 &
  180 &
  180 &
  12549 &
  232 \\
Repository Construction &
  - &
  58 &
  49 &
  41 &
  52 &
  57 &
  58 &
  45 &
  36 &
  3247 &
  341 \\ \midrule
\rowcolor{gray!15} \textbf{Total} &
  \multicolumn{1}{l}{} &
  \multicolumn{1}{l}{439} &
  \multicolumn{1}{l}{406} &
  \multicolumn{1}{l}{414} &
  423 &
  454 &
  455 &
  433 &
  403 &
  20742 &
  4617 \\ \bottomrule
\end{tabular}%
}
\caption{Statistic of \ourmethod{}. This table shows the number of examples and average token counts for each task and subtask across different programming languages.}

\label{tab:statistic}
\end{table*}

\subsubsection{Multi-file Editing}\label{sec:multi}
\paragraph{Task Settings}
The Multi-file Editing task evaluates the model's ability to add new functionalities across multiple files while preserving inter-file dependencies. Given a set of files \( \mathcal{C} = \{f_1, f_2, \ldots, f_m\} \), where each \( f_i \) is invoked by \( f_{i+1} \), and a functionality requirement \( R \), the task involves two scenarios:
\begin{itemize}
    \item \textit{In-place Edits}: Modify the existing files \( \mathcal{C} \) to incorporate \( R \), while maintaining inter-dependencies, resulting in a new set \( \mathcal{C}_{\text{new}} = \{f_1', f_2', \ldots, f_m'\} \).
    \item \textit{Expansion Edits}: Modify \( \mathcal{C} \) and create a new file \( f_{\text{new}} \) to integrate \( R \), resulting in an expanded set \( \mathcal{C}_{\text{new}} = \{f_1', f_2', \ldots, f_m'\} \cup \{f_{\text{new}}\} \), ensuring proper invocation relationships.
\end{itemize}
This approach reflects real-world scenarios where AI-driven code editors like Cursor\footnote{Cursor is an AI-powered code editor designed to assist developers. \url{https://www.cursor.com/}} help developers modify multiple files based on natural language requirements.

\paragraph{Sample Curation.}
We concatenate files from invocation chains of lengths 2, 3, and 4, then input these code snippets into GPT-4o. Using a step-by-step prompt, GPT-4o identifies cross-file interactions, proposes new features, and generates the corresponding modified code, including feature descriptions and explanations. Sample outputs for different invocation lengths and scenarios are provided in Appendix~\ref{prompt:multifile}.

\paragraph{Human in the Loop.}
We first apply an LLM-based filtering process to pre-select samples, ensuring basic quality and consistency with the task. GPT-4o evaluates samples based on predefined criteria, such as clarity of the feature description and logical implementation. Human annotators then manually review and correct the pre-selected samples, verifying the correctness of descriptions, logic, and code consistency. Annotators also ensure that errors in GPT-4o-generated code are corrected, curating gold labels for evaluation. Details about the process and our crowdsourcing can be found in Appendix~\ref{appendix:task1annotator} and Appendix~\ref{app:crowdsourcing}, respectively.

\subsection{Features of \ourmethod{}}

\paragraph{Statistic.} 
We present the statistic of \ourmethod{} in Table~\ref{tab:statistic}. We use the Qwen2.5-Coder~\cite{hui2024qwen25codertechnicalreport} tokenizer to compute the number of tokens. After the filtering and curation process, our test samples cover a total of 2,683 real-world repositories. Length distribution of \ourmethod{} is in Figure~\ref{fig:distribution}. More results about project topics are in Appendix~\ref{appendix:diversity}.

\paragraph{Future Extensions.}
\ourmethod{} currently supports 8 popular languages. In addition to releasing our test samples on GitHub, we have also open-sourced all the scripts used for crawling and curating these samples. Therefore, \ourmethod{} can potentially be extended to other languages and the community to use our tools to create more samples as needed. 
\begin{figure}[htbp]
\centering
\includegraphics[width=0.8\linewidth]{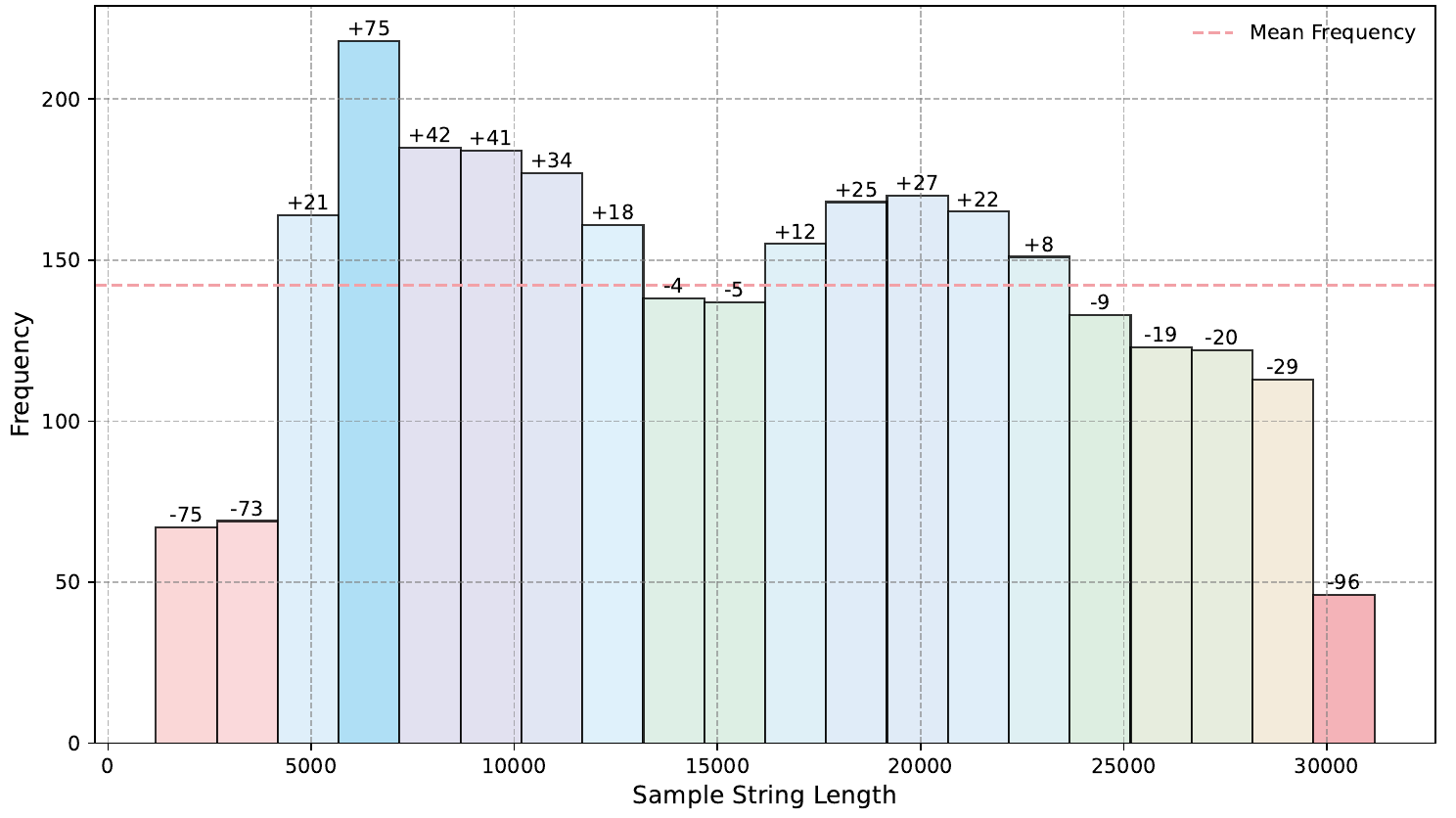}
\caption{Length Distribution of \ourmethod{}.}
\label{fig:distribution}
\end{figure}

\section{Experiments}
\begin{table*}[h]
\centering
\resizebox{1\textwidth}{!}{%
\begin{tabular}{cccccccccccccllcllcllcllcc}
\toprule
\multirow{2}{*}{\textbf{Models}} &
  \multicolumn{3}{c}{\textbf{C}} &
  \multicolumn{3}{c}{\textbf{C++}} &
  \multicolumn{3}{c}{\textbf{C\#}} &
  \multicolumn{3}{c}{\textbf{Python}} &
  \multicolumn{3}{c}{\textbf{Java}} &
  \multicolumn{3}{c}{\textbf{PHP}} &
  \multicolumn{3}{c}{\textbf{Typescript}} &
  \multicolumn{3}{c}{\textbf{Javascript}} &
  \textbf{Avg.} \\ \cline{2-26}\addlinespace[2pt] 
 &
  \textbf{DR} &
  \textbf{RC} &
  \textbf{ME} &
  \textbf{DR} &
  \textbf{RC} &
  \textbf{ME} &
  \textbf{DR} &
  \textbf{RC} &
  \textbf{ME} &
  \textbf{DR} &
  \textbf{RC} &
  \textbf{ME} &
  \multicolumn{1}{c}{\textbf{DR}} &
  \multicolumn{1}{c}{\textbf{RC}} &
  \textbf{ME} &
  \multicolumn{1}{c}{\textbf{DR}} &
  \multicolumn{1}{c}{\textbf{RC}} &
  \textbf{ME} &
  \multicolumn{1}{c}{\textbf{DR}} &
  \multicolumn{1}{c}{\textbf{RC}} &
  \textbf{ME} &
  \multicolumn{1}{c}{\textbf{DR}} &
  \multicolumn{1}{c}{\textbf{RC}} &
  \textbf{ME} &
  \textbf{-} \\ \midrule
\multicolumn{26}{c}{\textit{Open-Source Large Language Models (1.5B+)}} \\ \midrule
Yi-Coder-1.5B-Chat &
  7.41 &
  11.23 &
  0.45 &
  10.64 &
  9.62 &
  0.57 &
  4.84 &
  5.68 &
  0.01 &
  10.34 &
  8.80 &
  4.34 &
  0.00 &
  2.83 &
  0.35 &
  1.45 &
  14.52 &
  2.26 &
  8.47 &
  10.73 &
  0.50 &
  2.17 &
  4.93 &
  0.01 &
   5.09\\
OpenCoder-1.5B-Instruct &
  3.95 &
  9.03 &
  7.39 &
  1.59 &
  11.93 &
  13.24 &
  4.62 &
  9.58 &
  7.63 &
  4.55 &
  21.48 &
  16.66 &
  0.00 &
  6.98 &
  9.76 &
  2.17 &
  15.96 &
  7.70 &
  2.04 &
  10.98 &
  7.92 &
  3.66 &
  12.73 &
  8.17 &
  \resultthird{8.32} \\
Qwen2.5-Coder-3B &
  6.47 &
27.34 &
3.62 &
11.85 &
30.38 &
4.31 &
8.04 &
15.19 &
3.33 &
13.73 &
46.36 &
6.86 &
0.00 &
23.36 &
1.34 &
4.50 &
30.14 &
2.57 &
11.76 &
24.30 &
5.80 &
6.45 &
27.45 &
8.30 &
 \resultone{13.48}\\
Qwen2.5-Coder-3B-Instruct &
  4.46 &
5.23 &
4.64 &
9.44 &
8.26 &
4.51 &
5.59 &
5.72 &
4.66 &
8.93 &
14.20 &
9.95 &
10.00 &
2.15 &
6.49 &
7.22 &
7.37 &
5.98 &
5.59 &
5.52 &
3.55 &
9.55 &
6.51 &
10.61 & 
6.92\\
Granite-3b-code-instruct-128k &
  10.27 &
17.63 &
4.97 &
9.70 &
17.41 &
8.76 &
9.15 &
10.68 &
8.50 &
15.07 &
24.02 &
4.95 &
9.09 &
17.93 &
5.96 &
3.70 &
25.58 &
2.58 &
8.46 &
21.79 &
5.57 &
7.02 &
22.38 &
6.62 & 
\resulttwo{11.57}\\
Phi-3.5-mini-128k-instruct(3.82B) &
  8.22 &
3.43 &
1.31 &
13.38 &
15.84 &
2.01 &
13.48 &
0.81 &
1.01 &
11.19 &
9.01 &
3.01 &
5.11 &
2.38 &
1.01 &
17.24 &
13.06 &
0.77 &
11.51 &
5.76 &
1.01 &
17.05 &
5.98 &
1.42 & 
6.88 \\ \midrule
\multicolumn{26}{c}{\textit{Open-Source Large Language Models (7B+)}} \\ \midrule
Codestral-mamba(7B) &
  9.22 &
26.09 &
24.79 &
16.22 &
23.94 &
25.10 &
9.16 &
16.27 &
30.64 &
16.00 &
28.83 &
26.14 &
13.10 &
17.77 &
18.79 &
4.00 &
33.95 &
18.62 &
11.68 &
28.30 &
12.74 &
9.49 &
26.98 &
20.73 & 
19.52\\
CodeLlama-7b-Instruct-hf &
  8.00 &
18.21 &
13.23 &
11.81 &
18.01 &
13.25 &
6.25 &
10.26 &
12.67 &
7.63 &
30.43 &
18.42 &
5.74 &
16.91 &
12.67 &
4.00 &
27.67 &
11.43 &
9.52 &
20.41 &
11.32 &
6.96 &
21.03 &
10.12 & 
13.58\\
Yi-Coder-9B-Chat &
  15.85 &
36.33 &
8.62 &
19.14 &
48.06 &
10.55 &
11.46 &
15.67 &
6.02 &
13.67 &
28.89 &
21.04 &
13.60 &
25.12 &
13.39 &
12.50 &
35.51 &
9.78 &
12.98 &
35.37 &
7.95 &
11.45 &
32.29 &
6.25 & 
18.81\\
Granite-8b-code-instruct-128k &
  2.47 &
12.88 &
16.09 &
4.73 &
17.97 &
10.63 &
5.13 &
6.52 &
12.87 &
6.94 &
13.13 &
13.10 &
5.22 &
10.48 &
11.61 &
6.02 &
24.36 &
12.60 &
8.26 &
16.29 &
22.99 &
10.84 &
18.98 &
15.79 & 
11.91\\
Qwen2.5-Coder-7B &
  12.41 &
23.31 &
7.39 &
14.18 &
22.30 &
13.25 &
11.86 &
12.14 &
7.63 &
16.50 &
33.32 &
16.66 &
0.00 &
17.73 &
9.76 &
8.87 &
32.71 &
7.70 &
14.91 &
25.10 &
7.92 &
10.61 &
26.06 &
8.17 & 
15.02\\
Qwen2.5-Coder-7B-Instruct &
  16.85 &
20.30 &
17.58 &
26.16 &
21.90 &
19.18 &
20.11 &
7.69 &
14.87 &
28.81 &
19.37 &
19.23 &
18.99 &
19.80 &
13.18 &
22.01 &
19.71 &
15.31 &
30.87 &
15.97 &
18.34 &
24.12 &
19.02 &
15.29 & 
19.36\\
Phi-4(14B) &
  46.02 &
33.50 &
23.60 &
53.70 &
39.59 &
20.63 &
33.33 &
14.55 &
23.62 &
49.26 &
33.09 &
31.99 &
31.07 &
26.13 &
13.93 &
35.63 &
21.68 &
15.41 &
54.29 &
21.76 &
16.00 &
62.32 &
23.18 &
11.27 & 
\resulttwo{30.65}\\
CodeLlama-13b-Instruct-hf &
  6.41 &
20.09 &
18.23 &
6.76 &
21.02 &
21.25 &
9.09 &
11.74 &
16.67 &
11.88 &
29.04 &
16.42 &
7.79 &
18.64 &
18.67 &
7.02 &
28.97 &
8.42 &
7.10 &
21.91 &
13.32 &
10.37 &
25.66 &
13.21 & 
15.40 \\
Qwen2.5-Coder-14B &
  11.38 &
6.89 &
13.62 &
13.21 &
13.27 &
14.31 &
13.89 &
9.55 &
13.33 &
17.33 &
39.33 &
16.86 &
3.45 &
8.89 &
11.34 &
3.80 &
3.35 &
12.57 &
13.22 &
14.15 &
15.80 &
12.57 &
18.68 &
18.30 & 
13.30 \\ 
Qwen2.5-Coder-14B-Instruct &
  11.38 &
6.89 &
13.62 &
13.21 &
13.27 &
14.31 &
13.89 &
9.55 &
13.33 &
17.33 &
39.33 &
16.86 &
3.45 &
8.89 &
11.34 &
3.80 &
3.35 &
12.57 &
13.22 &
14.15 &
15.80 &
12.57 &
18.68 &
18.30 & 
\resultthird{26.70} \\
Codestral-2501(24B) &
  37.78 &
47.12 &
29.45 &
55.56 &
44.71 &
29.40 &
40.56 &
22.12 &
31.00 &
45.56 &
61.18 &
35.62 &
37.22 &
28.42 &
27.43 &
26.11 &
34.01 &
21.62 &
57.78 &
32.41 &
31.00 &
60.56 &
33.69 &
33.50 & 
\resultone{37.66} \\ \midrule

\multicolumn{26}{c}{\textit{Open-Source Large Language Models (32B+)}} \\ \midrule
CodeLlama-34b-Instruct-hf &
  8.77 &
24.59 &
23.00 &
5.62 &
23.89 &
31.00 &
1.44 &
11.39 &
27.00 &
10.56 &
33.01 &
31.28 &
2.63 &
20.58 &
26.32 &
2.37 &
31.53 &
19.67 &
3.79 &
23.83 &
17.32 &
3.55 &
24.02 &
21.23 & 
17.85\\
Qwen2.5-Coder-32B-Instruct &
  52.22 &
49.40 &
30.29 &
70.00 &
58.27 &
36.67 &
42.78 &
24.06 &
41.00 &
61.11 &
46.93 &
37.12 &
43.89 &
33.04 &
35.92 &
50.56 &
37.47 &
14.71 &
63.13 &
41.47 &
44.80 &
57.22 &
35.96 &
40.36 & 
\resulttwo{43.68}\\
Qwen2.5-Coder-32B &
  15.88 &
17.26 &
23.50 &
25.77 &
18.20 &
33.42 &
18.89 &
11.37 &
25.86 &
21.66 &
23.20 &
23.11 &
20.51 &
13.03 &
17.00 &
14.56 &
26.63 &
20.45 &
26.14 &
19.37 &
15.35 &
24.40 &
26.30 &
20.71 & 
20.94\\
Llama-3.3-70b-instruct &
  17.78 &
10.00 &
9.49 &
34.64 &
12.00 &
14.88 &
16.29 &
30.02 &
11.38 &
9.44 &
22.10 &
18.54 &
2.22 &
12.30 &
18.52 &
17.22 &
12.00 &
8.94 &
13.89 &
10.09 &
6.19 &
39.66 &
18.12 &
5.08 & 
15.45\\
Qwen-2.5-72b-instruct &
  43.89 &
36.33 &
15.72 &
67.22 &
48.61 &
21.60 &
35.56 &
16.32 &
37.21 &
51.11 &
41.33 &
22.93 &
40.56 &
15.47 &
33.06 &
43.33 &
27.53 &
32.18 &
65.56 &
29.61 &
19.26 &
61.67 &
29.83 &
17.83 & 
\resultthird{35.57}\\
DeepSeek-Coder-V2-Lite-Instruct(21/236B) &
  8.94 &
21.48 &
24.87 &
11.67 &
25.29 &
28.69 &
7.78 &
12.49 &
25.86 &
11.67 &
20.45 &
24.61 &
16.67 &
12.02 &
39.97 &
12.78 &
19.96 &
11.67 &
12.43 &
25.15 &
39.59 &
14.86 &
19.50 &
35.59 & 
20.17\\
DeepSeek-V3(37/671B) &
  68.54 &
64.25 &
36.97 &
83.05 &
63.15 &
47.46 &
54.75 &
26.85 &
41.22 &
63.48 &
67.89 &
37.99 &
48.31 &
33.51 &
44.32 &
67.05 &
51.58 &
40.54 &
83.33 &
53.65 &
43.60 &
88.83 &
53.13 &
40.11 & 
\resultone{54.31} \\ \midrule
\multicolumn{26}{c}{\textit{Closed-Source Large Language Models (API)}} \\ \midrule
Claude-3.5-sonnet-20241022 &
  52.78 &
63.12 &
42.00 &
55.56 &
66.00 &
57.00 &
30.00 &
27.93 &
43.75 &
25.56 &
61.95 &
55.83 &
9.44 &
30.81 &
47.67 &
48.33 &
60.92 &
44.33 &
53.89 &
57.62 &
46.33 &
62.78 &
57.38 &
42.12 & 
\resultone{47.63}\\
GPT-4o-mini-2024-07-18 &
  50.56 &
41.90 &
39.55 &
60.00 &
49.59 &
45.17 &
30.00 &
13.56 &
41.06 &
51.67 &
41.43 &
38.37 &
37.78 &
31.16 &
44.93 &
47.19 &
25.71 &
25.73 &
51.67 &
28.60 &
42.87 &
60.00 &
26.07 &
39.78 & 
\resultthird{40.18}\\
 Palm-2-codechat-bison &
  8.89 &
22.49 &
12.37 &
11.17 &
33.29 &
16.80 &
10.98 &
3.03 &
20.83 &
9.60 &
10.61 &
25.46 &
9.44 &
2.69 &
14.11 &
9.71 &
19.62 &
18.82 &
9.77 &
16.25 &
16.15 &
9.66 &
13.82 &
16.05 & 
14.23\\ 
Gemini-1.5-flash-latest &
52.78 &
52.17 &
36.42 &
58.33 &
56.19 &
44.49 &
36.67 &
22.10 &
39.88 &
58.89 &
27.49 &
36.32 &
37.78 &
13.10 &
47.05 &
35.00 &
29.42 &
33.24 &
58.89 &
44.14 &
57.18 &
57.78 &
35.90 &
46.88 &
\resulttwo{42.42} \\
\bottomrule
\end{tabular}%
}
\caption{Results of different models on the 
\ourmethod{}. We utilize \resultone{green}(1st) \resulttwo{blue}(2nd) \resultthird{yellow}(3rd) to distinguish the top three results within different sizes.}
\label{tab:my-table}
\end{table*}

\begin{figure*}[!htbp]
    \centering
    \adjustbox{center}{\includegraphics[width=1\textwidth]{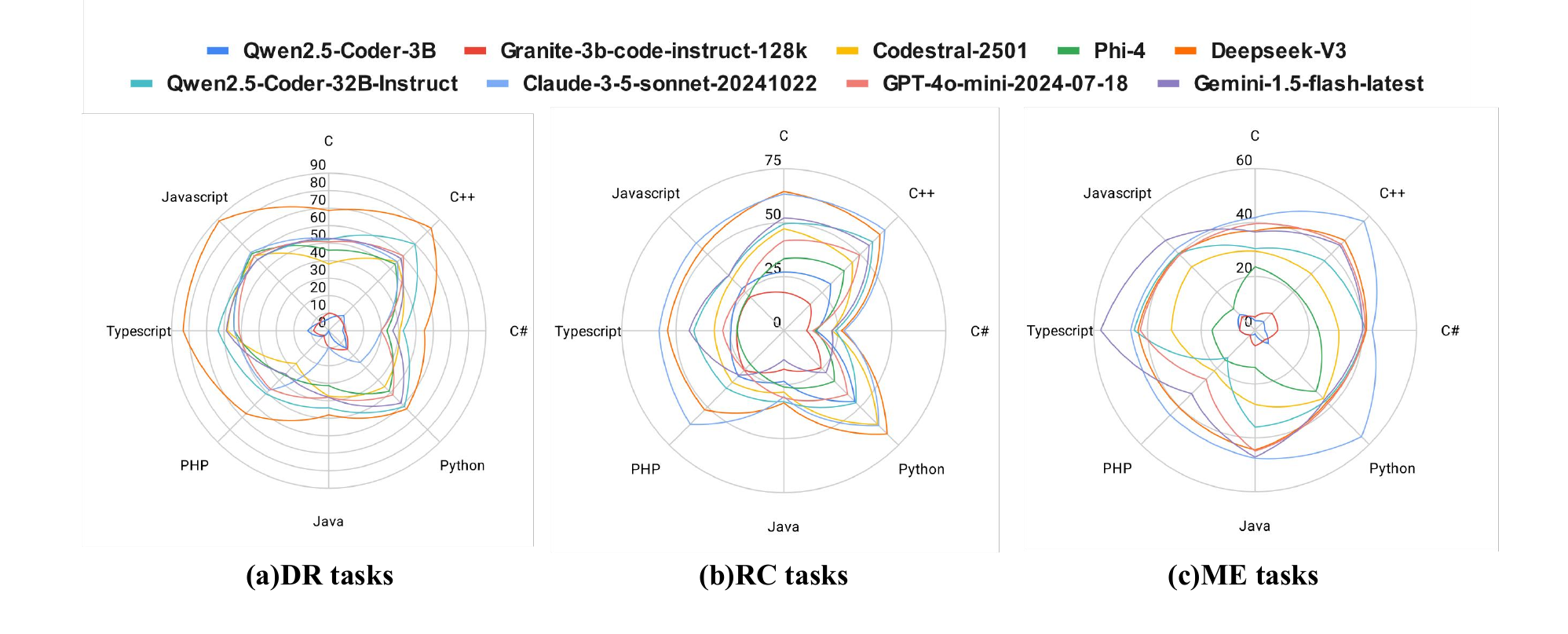}}
    \caption{Results of different tasks in \ourmethod{}. The radar charts show the performance of various models across Dependency Recognition(a), Repository Construction(b), and Multi-file Editing(c) tasks. Each line represents a different model, with performance measured for different programming languages.}
    \label{fig:rader3}
\end{figure*}
\subsection{Models}
We evaluate over 25 models, ranging from 1.5B to 670B+ parameters, including both open-source large language models and closed-source general LLMs. For general models, we test GPT-4o-mini~\citep{gpt4}, Claude series~\cite{claude}, Phi series~\citep{abdin2024phi3technicalreporthighly, abdin2024phi4technicalreport}, PaLM2-CodeChat~\cite{anil2023palm2technicalreport}, Llama3.3\footnote{\url{https://ai.meta.com/blog/meta-llama-3/}}, and Qwen2.5~\cite{qwen2025qwen25technicalreport}. For code models, we evaluate CodeLlama~\cite{code_llama}, OpenCoder~\cite{opencoder}, Qwen-Coder~\citep{qwen25coder}, DeepSeekCoder~\citep{deepseek_coder}, CodeStral~\citep{codestral}, Yi-Coder\footnote{\url{https://huggingface.co/01-ai/Yi-Coder-9B}}, and Granite-Coder~\cite{granite_coder}.

\subsection{Implementation Details}
Our evaluation framework is built on the Transformers library~\cite{wolf2020huggingfacestransformersstateoftheartnatural}. We use a one-shot prompt for the Dependency Recognition and Repository Construction tasks, while the Multi-File Editing task is performed in a zero-shot setting, without additional training. All experiments are run on 16 NVIDIA A800-SXM4-80GB GPUs with the same hyperparameters for code generation across all models. Detailed settings and prompts are provided in Appendix~\ref{prompts:inference}.

\subsection{Evaluation Metrics}

\paragraph{Dependency Recognition.}  
For the Dependency Recognition task, we calculate the \textbf{Exact Match Rate} (EMR), which measures the proportion of instances where the predicted dependency chain exactly matches the ground truth:
\begin{equation}
\text{EMR} \stackrel{\triangle}{=} \frac{|\{ i \mid EM_i = 1 \}|}{N}
\end{equation}
where
\begin{equation}
EM_i =
\begin{cases}
    \frac{\sum_{j} \mathbb{1}(p_j = g_j)}{|P_i|}, & \text{if } |P_i| > 0 \\
    0, & \text{otherwise}
\end{cases}
\end{equation}
Here, \( P_i \) and \( G_i \) denote the predicted and ground-truth dependency sets, and \( \mathbb{1}(\cdot) \) is an indicator function, capturing the proportion of fully correct predictions.

\paragraph{Repository Construction.}  
For Repository Construction, we evaluate structural similarity by comparing the predicted and ground-truth dependency graphs. Given a set of predicted file invocation chains, we construct a directed graph \( G_p = (V_p, E_p) \) and similarly for the ground-truth graph \( G_g = (V_g, E_g) \). We then compute precision, recall, and F1-score for both nodes and edges:
\begin{equation}
\text{F1} = \frac{2 \times \text{Precision} \times \text{Recall}}{\text{Precision} + \text{Recall}}
\end{equation}
Finally, we combine node and edge F1-scores (\emph{i.e.} $\text{F1}_\text{n}$ and $\text{F1}_\text{e}$) as follows:
\begin{equation}
\text{F1}_\text{com} = w_1 \times \text{F1}_\text{n} + w_2 \times \text{F1}_\text{e},
\end{equation}
where $w_1$ and $w_2$ are empirically set to $0.15$ and $0.85$, respectively.

\paragraph{Multi-file Editing.}

Inspired by FairEval~\cite{faireval} to reduce bias in LLMs as judge models, we prompt the large language model (LLM), denoted as $\text{LLM}(\cdot)$, to evaluate the correctness and completeness of multi-file edits. The LLM compares the generated code with the corrected version provided by human annotators in section~\ref{sec:multi}. We define the following metrics: Correctness ($C$): Measures how accurately the files meet the expected functionality. Purpose Alignment ($PA$): Assesses how well the code aligns with its intended purpose. Functionality Accuracy ($FA$): Evaluates the accuracy of the functionality across files. Functionality Completeness ($FC$): Checks if all required aspects of functionality are addressed. Code Quality ($CQ$): Reflects the overall quality of the code. The detailed prompts are in Appendix~\ref{prompts:evaluation}. The combined score is calculated as:
\begin{equation}
\begin{split}
    \text{Score} = &\ \lambda_1 \cdot \text{LLM(C)} + \lambda_2 \cdot \text{LLM(PA)} \\
    &+ \lambda_3 \cdot \text{LLM(FA)} + \lambda_4 \cdot \text{LLM(FC)} \\
    &+ \lambda_5 \cdot \text{LLM(CQ)},
\end{split}
\end{equation}
where the weighting parameters are empirically set as follows: \(\lambda_1 = \lambda_2 = 0.25\), \(\lambda_3 = \lambda_4 = 0.20\), and \(\lambda_5 = 0.10\).

\subsection{Main Results}
In the following discussions, we will refer to specific tasks using the following abbreviations: ME (Multi-file Editing), DR (Dependency Recognition), and RC (Repository Construction).
\subsubsection{Overall Evaluation}
\paragraph{Model Size \& Type and Performance.}  
Larger models generally outperform smaller ones, as seen in open-source architectures like \texttt{Codellama-34B-Instruct} and \texttt{Qwen-Coder-32B}, which consistently show better performance than their smaller counterparts. This suggests that increasing model capacity enhances the ability to manage complex dependency relationships in code repositories. Interestingly, \texttt{Qwen-Coder-32B} surpasses larger general models like \texttt{Llama-3.3-70b-instruct} and \texttt{Qwen-2.5-72b-instruct}, highlighting that domain-specific training can be more effective than simply scaling model size. This implies that targeted pretraining on high-quality code datasets and instruction tuning for software engineering tasks are key to improving performance.

\paragraph{Task-Specific.} 
Larger models show clear advantages in DR and RC tasks. For instance, \texttt{DeepSeek-V3(37/671B)} excels in DR, demonstrating strong ability to understand complex inter-file dependencies. However, Multi-file Editing remains a more intricate task, even larger models face challenges in maintaining consistency and accuracy across multiple files
\paragraph{Language-Specific.} 
Closed-source models generally show more stable performance across languages in the ME task. Performance also varies by language, with Python and JavaScript yielding better results, particularly in Dependency Recognition and Repository Construction tasks. In contrast, languages like C\# and Java are more challenging. Models perform better in languages with static, modular structures, while languages with complex interdependencies or dynamic features pose more difficulty for multi-file editing and dependency recognition.
\paragraph{Closed-source vs. Open-source Models.} 
On average, closed-source models outperform most open-source ones. However, \texttt{DeepSeek-V3} surpasses \texttt{Claude-3.5-sonnet-20241022}, showing that top-tier open-source models remain competitive. While models like \texttt{Qwen2.5-Coder-32B} and \texttt{DeepSeek-V3} excel in simpler tasks like DR, they struggle with more complex ones like ME. In contrast, closed-source models, especially \texttt{Claude-3.5-sonnet-20241022}, perform better in RC and ME, indicating stronger reasoning and generalization abilities for multi-file editing and repository understanding.

\subsubsection{Takeaways from Task-specific Evaluation}

\noindent \textbf{Dependency Relationships are Crucial for ME.} Models strong in Dependency Recognition excel in Multi-file Editing. This correlation shows that capturing and modifying dependencies is key to accurate, consistent changes across large codebases.


\noindent \textbf{LLMs Struggle with Directory Structure.} LLMs may lack the prior knowledge needed to effectively organize hierarchical project structures, which is reflected in the generally lower RC scores compared to DR. To improve performance, incorporating more structured code repositories into the training data could help LLMs learn better patterns for organizing directory structures.

\noindent \textbf{Cross-File Modifications Remain Challenging.} Even closed-source models struggle with cross-file modifications in ME, requiring better consistency and a global view across files. Improving training with real-world ME examples could help.







\section{Analysis} \label{analysis}

\subsection{Multilingual Analysis}
Figure~\ref{fig:rader3} shows the performance of various models across multiple programming languages for three tasks. We selected the top two models for each parameter range. Key insights include as follows.

\noindent \textbf{LLMs struggle with statically typed languages.} In Dependency Recognition and Repository Construction, languages like Python, JavaScript, and TypeScript perform better due to their clear modular structures and explicit imports. In contrast, PHP and C\# present challenges with flexible file inclusion and complex project files, respectively.
\noindent \textbf{Intricate interdependencies challenge multi-file editing (ME).} Models perform worse in C and C++, where complex cross-file modifications are needed, like pointer tracking and header updates. Languages such as Python and Java, with structured class designs, perform better in ME.

 \subsection{Instruction Following}
\begin{figure}[t]
    \centering
    \includegraphics[width=\columnwidth]{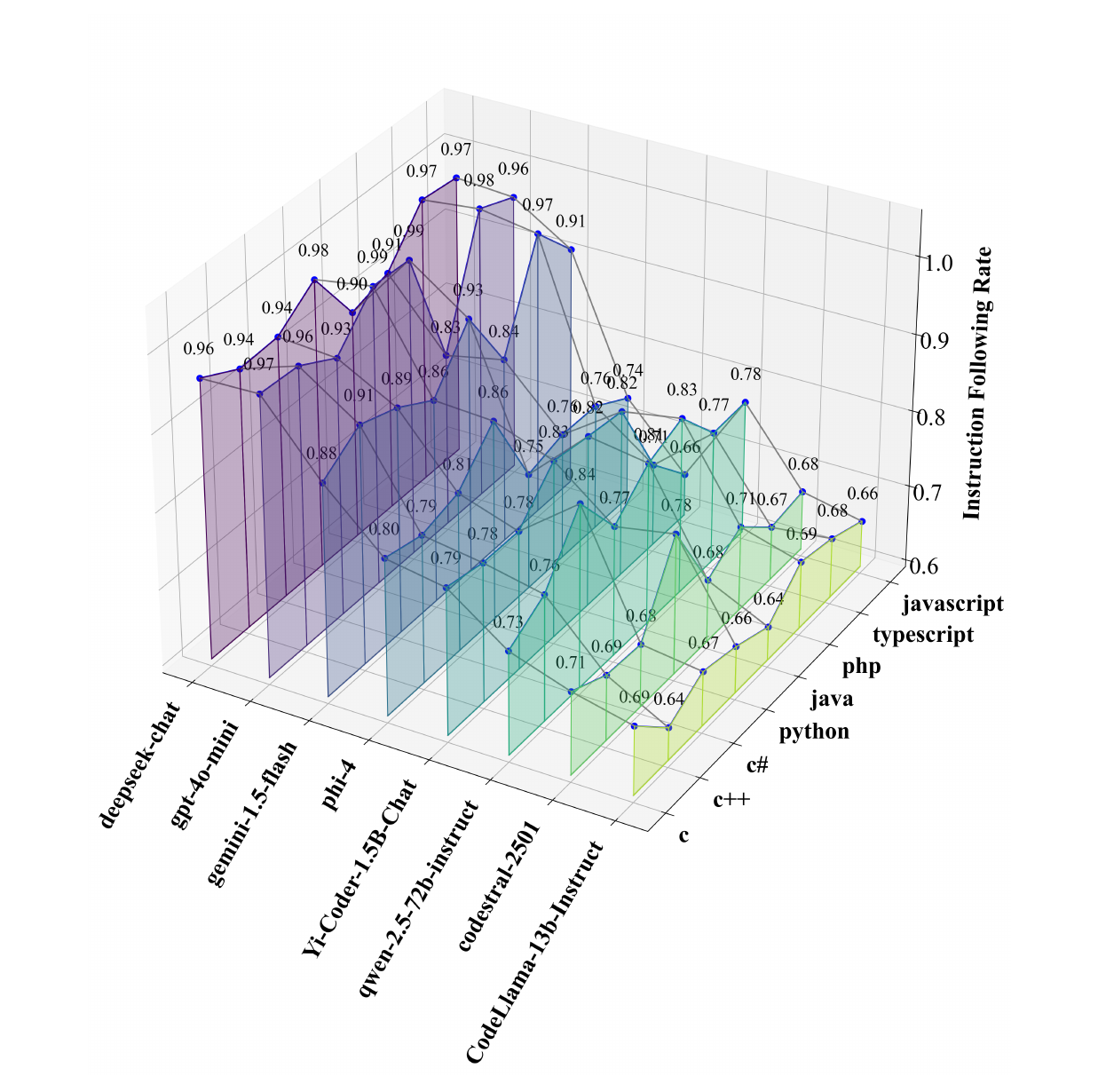}
    \caption{Instruction-following performance covering 8 different languages.}
    \label{fig:instruction1}
\end{figure}

\begin{figure}[htbp]
    \centering
    \includegraphics[width=\columnwidth]{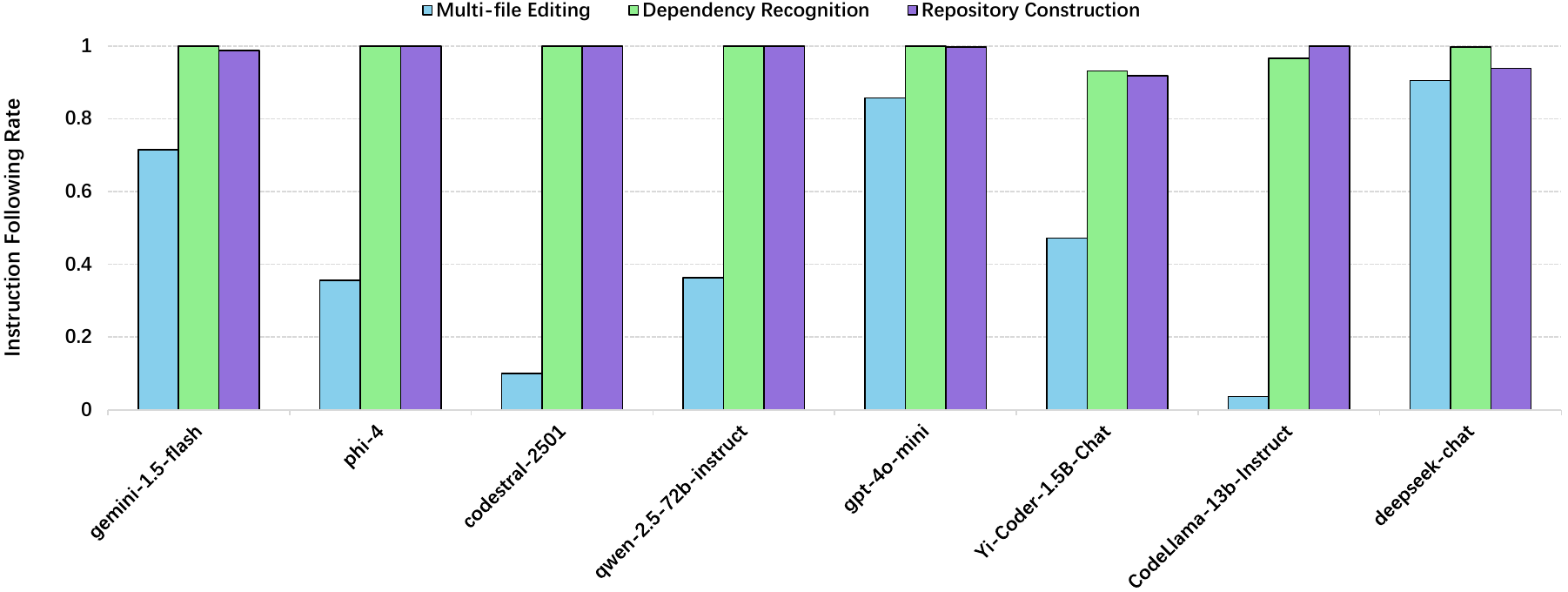}
    \caption{Instruction-following performance covering 3 different tasks.}
    \label{fig:instruction2}
    \vspace{-5mm}
\end{figure}

\noindent \textbf{Language-wise Analysis.}
In Figure~\ref{fig:instruction1}. \texttt{deepseek-chat} is the strongest model overall, demonstrating its strong adaptability across various programming languages. More complex languages, such as C++ and PHP, pose challenges for all models. Most models show weaker instruction-following abilities in these languages, indicating poor understanding and execution capabilities. In contrast, languages like JavaScript and Python are relatively easier to process, resulting in higher instruction-following rates for these languages.

\noindent \textbf{Task-wise Analysis.} In Figure~\ref{fig:instruction2}. At the task level, \texttt{deepseek-chat} significantly outperforms other models, showcasing its efficient handling of multiple tasks. Most models perform well on the \texttt{Repository Construction} and \texttt{Dependency Recognition} tasks, which are relatively easier as they mainly focus on file relationships and dependencies. On the other hand, the \texttt{Multi-file Editing} task is more challenging, requiring models to simultaneously understand modifications across multiple files and their collaboration, which leads to poorer performance for some models.

\section{Conclusion}

We introduce \ourmethod{}, a benchmark designed to evaluate the repository-level dependency understanding of LLMs. Built on a diverse set of 15,576 real-world repositories, the benchmark covers three key tasks: Dependency Recognition, Repository Construction, and Multi-file Editing, across eight programming languages. Our evaluation of over 25 LLMs reveals significant performance gaps, highlighting the difficulties models face when managing complex code repositories. These insights point to the need for LLMs to improve their handling of dependencies, project structures, and multi-file modifications. Our work sets the stage for future advancements in enhancing the repository-level reasoning capabilities of LLMs.

\clearpage
\section{Limitations}
\label{app:limitation}
\paragraph{Not Enough Languages.}  
While \ourmethod{} evaluates models across 8 programming languages, this scope is limited and may not fully capture the performance of models in other popular or emerging languages. Expanding the range of supported languages will provide a more comprehensive evaluation of LLMs' capabilities across diverse coding environments.

\paragraph{Not Enough Tasks.}  
Currently, the benchmark focuses on three core tasks: Dependency Recognition, Repository Construction, and Multi-file Editing. While these tasks are crucial, they do not encompass all the challenges faced in real-world software development. Including additional tasks such as debugging, code refactoring, or performance optimization would offer a more complete assessment of LLMs in software engineering contexts.

\paragraph{Not Enough Models.}  
Although we evaluate over 25 models, this number is not sufficient to represent the full range of LLMs available, particularly as new models continue to emerge. Expanding the model pool would provide deeper insights into the performance of various model architectures and sizes, ensuring a more robust evaluation.

\paragraph{Future Updates.}  
As the field of large language models evolves, \ourmethod{} will be updated to include more languages, tasks, and models. This will help ensure the benchmark remains relevant and continues to provide valuable insights into the growing capabilities of LLMs in handling complex software engineering tasks.

\bibliography{custom}

\clearpage
\appendix
\onecolumn

\section{Import Statements}\label{appendix:Import statement}

\begin{mybox}
    \textbf{Python}\
    \textbf{Basic import(and rename)}: Use the "import" keyword followed by the module name  or "as" keyword to give it an alias, like "import xx","import xx as xxx".
    
    \textbf{Import specific content from a module(and rename)}: Use the "from" keyword followed by the module name and "import" keyword (and "as" keyword), such as "from a import b as c".
    
    \textbf{Import multiple functions}: use a comma-separated list within the import statement for importing multiple functions from a module or package,like "from mymodule import function1, function2".
 
     \textbf{Absolute references}: An absolute reference specifies the complete path to a resource from the root directory. For instance, "from mypackage.mymodule","import myfunction".
     
    \textbf{Relative references}: A relative reference specifies a path starting from the current location in the directory structure. For example, "from . import sibling\_module".
\end{mybox}

\begin{mybox}
    \textbf{C}\
    \textbf{Basic include}: Use the \#include directive followed by the header file name. Standard library headers are enclosed in angle brackets < >, while user-defined headers are enclosed in double quotes " ". For example, \#include <xx.h> or \#include "xx.h".

    \textbf{Include specific content from a header file(and rename)}: C can use typedef or \# define to rename data types or functions. For example, typedef old\_type new\_type; or \#define new\_func old\_func.
    
    \textbf{Absolute references}: An absolute reference specifies the complete path to a resource from the root directory. In C, this is typically used for specifying the full path to a header file or source file. For example, \#include "/usr/include/myheader.h".

    \textbf{Relative references}: A relative reference specifies a path starting from the current location in the directory structure. For example, \#include "./myheader.h" includes a header file from the current directory, or \#include "../myheader.h" includes a header file from the parent directory.
\end{mybox}

\begin{mybox}
    \textbf{C++}\
    \textbf{Basic include}: Use the \#include directive followed by the header file name. Standard library headers are enclosed in angle brackets < >, while user-defined headers are enclosed in double quotes " ". For example, \#include <xx.h> or \#include "xx.h".

    \textbf{Include specific content from a header file(and rename)}: C++ allows you to include specific parts of a header file by using namespace to refer to certain namespaces or classes. You can also rename or alias types or functions using using directives, like "using namespace std" or "using std::vector".

    \textbf{Absolute references}: An absolute reference specifies the complete path to a resource from the root directory. In C++, this is typically used for specifying the full path to a header file or source file. For example, \#include "/usr/include/myheader.h".

    \textbf{Relative references}: A relative reference specifies a path starting from the current location in the directory structure. For example, \#include "./myheader.h" includes a header file from the current directory, or \#include "../myheader.h" includes a header file from the parent directory.
\end{mybox}

\begin{mybox}
    \textbf{C\#}\
    \textbf{Basic using)}: In C\#, you use the "using" keyword to import namespaces, which allow access to classes, structs, and other members within those namespaces. For example, "using System"; allows access to the classes in the "System" namespace. 

    \textbf{Import specific content from a namespace}: C\# can import specific types (such as classes or methods) from a namespace by "using" followed by the type name. For instance, "using System.Console". 

    \textbf{Absolute references}: In C\#, absolute references are typically used for referencing fully qualified type names (including the namespace). For example, "System.Console.xx".
\end{mybox}

\begin{mybox}
    \textbf{PHP}\
    \textbf{Basic include}: Use the "include" or "require" keyword followed by the path to the file you wish to include. You can also use include\_once or require\_once to ensure the file is included only once. For example, "include 'myfile.php'" or "require 'myfile.php'".

    \textbf{Include specific content from a file}: PHP can include an entire file or use functions and classes defined in the included file. To rename or alias functions, classes, or variables, you would use PHP’s "use" keyword in the context of namespaces. For example, "use MyNamespace".

    \textbf{Absolute references}: An absolute reference specifies the full path from the root directory. In PHP, this would be used to include files from a specific path on the server, such as: "include '/xx/xx/xx.php'".

    \textbf{Relative references}: A relative reference specifies a path relative to the current directory. For example, "include './xx.php'" or "include '../xx.php'".
\end{mybox}

\begin{mybox}
    \textbf{JAVA}\
    \textbf{Basic import}: In Java, the "import" keyword is used to include classes or entire packages from other files. By default, Java does not support aliasing in imports like some other languages. To import a class, use "import package.ClassName".

    \textbf{Import all classes from a package}: Java allows importing all the classes from a package by using the * wildcard. For example, "import java.util.*".

    \textbf{Absolute references}: An absolute reference in Java specifies the full path to a class, starting from the root of the classpath. For example, "import xx.xx.MyClass".

    \textbf{Static imports}: Java allows importing static members of a class, such as methods or constants, using the import static keyword. For example, import static java.lang.Math.*; allows access to static methods like Math.sqrt() without needing to prefix them with Math.
\end{mybox}

\begin{mybox}
    \textbf{JavaScript}\
    \textbf{Basic import(and rename)}: Use the import keyword to bring in modules or specific parts of modules. The general syntax is import <module> from '<module-name>', where <module> is the exported entity. To rename an import, the as keyword can be used, like import { originalName as alias } from '<module-name>'.

    \textbf{Import specific content from a module(and rename)}: JavaScript allows importing specific functions, objects, or values from a module. This can be done by using the {} syntax. Additionally, renaming is possible with the as keyword. For example, import { func1 as f1, func2 } from '<module-name>' imports func1 as f1 and func2 as-is.

    \textbf{Import multiple items from a module}: Multiple exports from the same module can be imported in a single statement. These are separated by commas inside curly braces. For example, import { func1, func2, constant } from '<module-name>'.

    \textbf{Absolute references}: An absolute reference in JavaScript specifies the complete URL or file path to the module or resource. This often starts from the root directory or the full URL. For example, import { func } from '/modules/myModule.js' or import { func } from 'https://example.com/myModule.js'.

    \textbf{Relative references}: A relative reference in JavaScript refers to the file path starting from the current location in the directory structure. This can be done using ./ for the current directory or ../ for the parent directory. For example, import { func } from './myModule.js' imports from the current directory, while import { func } from '../myModule.js' imports from the parent directory.
\end{mybox}

\begin{mybox}
    \textbf{TypeScript}\
    \textbf{Basic import(and rename)}: Use the import keyword to bring in modules or specific components from modules. The syntax is import <module> from '<module-name>', where <module> refers to the default export of the module. To rename an import, the as keyword is used, such as import { originalName as alias } from '<module-name>'.

    \textbf{Import specific content from a module(and rename)}: TypeScript supports importing specific parts of a module using curly braces. The as keyword can also be used to rename the imported elements. For example, import { func1 as f1, func2 } from '<module-name>' imports func1 as f1 and func2 with its original name.

    \textbf{Import multiple items from a module}: Multiple items from the same module can be imported by listing them inside curly braces, separated by commas. For instance, import { func1, func2, variable } from '<module-name>' imports multiple functions and variables from the module.

    \textbf{Absolute references}: An absolute reference in TypeScript specifies the full path to a module, either from the root directory or via a full URL. For example, "import { myFunction } from '/xx/myModule'" or "import { myFunction } from 'https://xx/myModule'".

    \textbf{Relative references}: A relative reference specifies the file path starting from the current directory. For example, "import { myFunction } from './myModule'" or "import { myFunction } from '../myModule'".
    
\end{mybox}

\section{Repository Collection}\label{appendix:filter}
\subsection{Repo Filter Criteria}

We collect public GitHub repositories created before December 16, 2024, and focus on eight specific programming languages. To streamline data processing, we apply several filtering criteria.

First, we exclude files where the average line length exceeds 100 characters or the maximum line length surpasses 1000 characters. Additionally, files with fewer than 25\% alphabetic characters are removed.

At the repository level, we enforce the following main conditions. (1) Repositories must fall within the specified creation timeframe and must not be forks to ensure originality. (2) They must be primarily written in one of the eight target programming languages listed in Table \ref{tab:statistic}. (3) They must meet size and popularity constraints: repositories should be smaller than 1MB for manageability, and they must have at least three stars to reflect a minimum level of community interest. (4) Repositories must have at least one commit in the last six months before the collection date to ensure they are not abandoned.
(5) Repositories must have at least one open or closed issue or pull request to indicate some level of engagement from developers.
(6)  For repositories using dependency managers (e.g., package.json for JavaScript, requirements.txt for Python), we check that these files are present and not empty to ensure the repository is functional.
(7) Repositories must contain a README file with at least 100 words to ensure they provide sufficient context for analysis.
(8) To maintain consistency, we exclude repositories that mix multiple programming languages beyond a small threshold (e.g., more than 10\% of files in a secondary language).

\subsection{README Filter Criteria}
To ensure that the README file effectively serves its purpose by providing structured and comprehensive information, we apply the following filtering criteria: (1) Structured Formatting: The README must include appropriate titles and headings to organize content, making it easier for readers to navigate.
(2) Content Sufficiency: The document should meet a minimum character threshold to ensure it provides a meaningful overview of the project rather than being overly brief.
(3) Setup Instructions: The README must contain setup-related keywords such as install, build, setup, download, compile, train, or run, ensuring that it provides essential instructions for setting up and using the project.
(4) Project Structure References: It should explicitly reference specific files or directories within the repository to help users understand the project’s organization and locate key resources quickly.
(5) License Information: The presence of a license section or a direct reference to a LICENSE file is required to clarify usage permissions and distribution rights.

\section{Human Review Criteria}
\subsection{Repository Construction}\label{appendix:task4annotator}
Human annotators perform the following criteria to select the final set of samples:

First, the repository description should provide a brief and clear summary of the purpose and functionality of the repository. It should be logically coherent, free from contradictions, and align with the functional components of the repository. In addition, the description should include all important features and functionalities of the repository, ensuring a comprehensive overview.
Next, the functional description for each file should clearly outline the core features and capabilities of the file. It should be accurate and brief, reflecting the actual code behavior of the file while adhering to the conventions and terminology of the repository. 

Finally, the relationships between files, functions, and modules should be accurately represented in the generated descriptions. The dependencies of each file and the interactions within the repository must be clearly conveyed, ensuring that the dependencies reflect the actual structure of the repository.

\subsection{Multi-file Editing}\label{appendix:task1annotator}
After the LLM-based filtering step, pre-selected samples are passed to human annotators for manual verification. This process is designed to ensure accuracy and consistency across multiple files. 

First, annotators check whether the new functionalities are described in a way that is consistent with the code's behavior and the intended programming logic. Next, They should confirm that the changes align with the existing codebase are semantically correct and do not introduce errors or bugs. Modifications must be coherent not only within the specific file but throughout the project, maintaining consistency in terms of structure, style, and performance. Then, annotators should cross-check that all modified files reflect the appropriate changes, ensuring that the changes in one file do not conflict with others. Finally, annotators manually correct any errors, such as syntax issues or logical inconsistencies, refining the code to produce gold labels that serve as the standard for subsequent evaluations.

\section{Diversity of Data}\label{appendix:diversity}
\begin{figure}[htbp]
    \centering
    \adjustbox{center}{\includegraphics[width=0.5\linewidth]{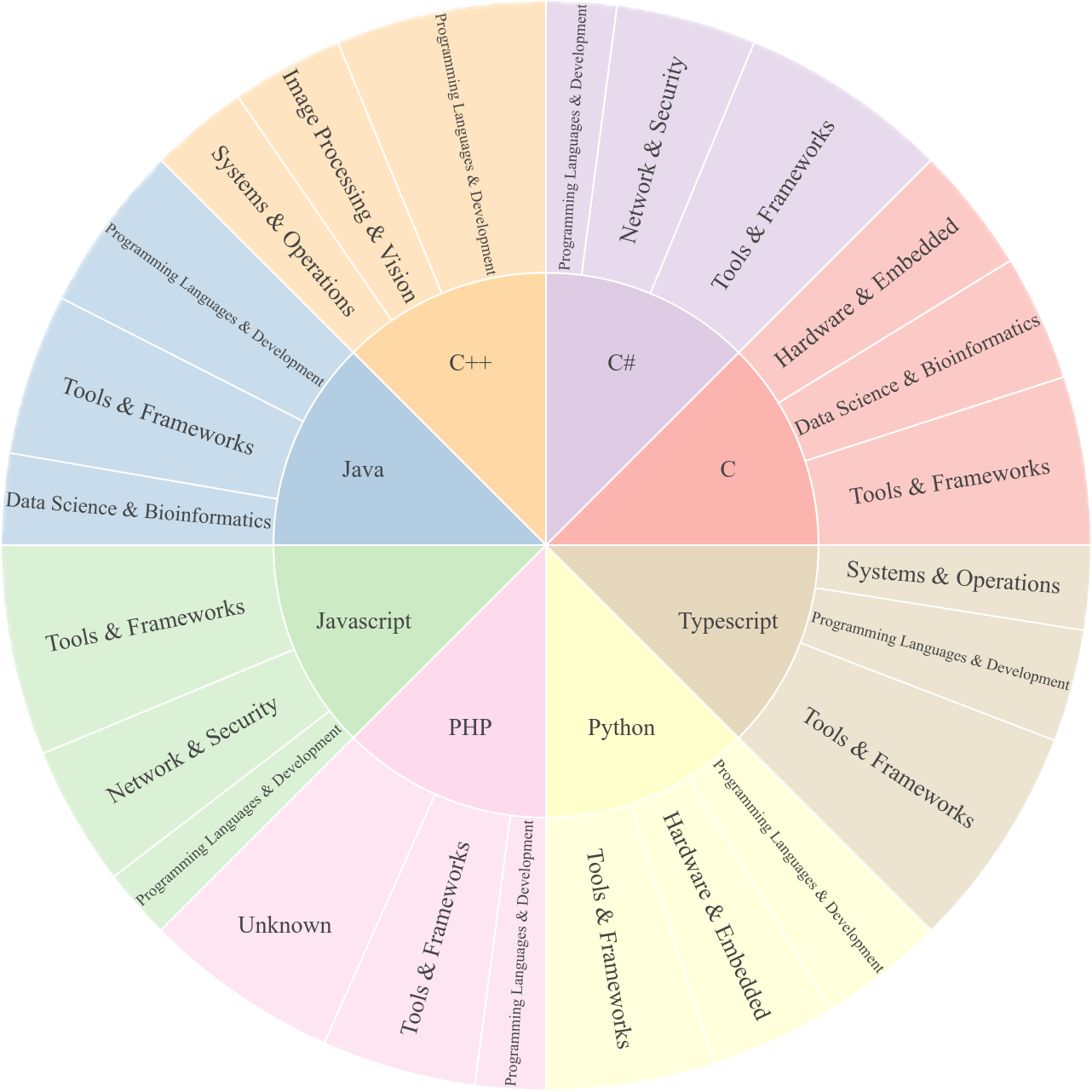}}
    \caption{Topic distribution. Different programming languages and frameworks into various fields, such as Development, Tools, Systems, Data Science, and more.}
    \label{fig:taskcase}
\end{figure}

\section{Prompt Templates}

\subsection{
Curating Prompts for Dependency Recognition Data}\label{prompt:datacuration}

\begin{tcolorbox}[breakable, colback=blue!5!white, colframe=blue!50!white, title= Data Curation for Dependency Recognition]
You are an AI assistant that summarizes file functionality. Your task:\
1. Read the file content provided by the user.\
2. Analyze and understand the file's main purpose.\
3. Summarize the core functionality in 1-2 concise sentences.\
4. Provide only the summary, without any additional information or commentary.
\label{Dependency_Recognition_prompt}
\end{tcolorbox}

\begin{tcolorbox}[breakable, colback=blue!5!white, colframe=blue!50!white, title= Data Curation for Dependency Recognition]
You are an AI assistant that summarizes project functionality. Your task:\
1. Read the README content provided by the user.\
2. Analyze and understand the project's main purpose and features.\
3. Summarize the core functionality of the project in 2-3 concise sentences.\
4. Provide the summary in JSON format with two keys: "description" and "function".\
5. The "description" should briefly describe what the project is.\
6. The "function" should summarize the main features or capabilities of the project.

Example output:

\noindent\begin{minipage}{\columnwidth}
\begin{lstlisting}[language=json]
{
    "description": "An AI-powered game solver for 2048",
    "function": "Utilizes expectimax optimization with efficient bitboard representation to play 2048. Can control browser-based game and provides both command-line and browser-control versions."
}
\end{lstlisting}
\end{minipage}
\label{Dependency_Recognition_prompt}
\end{tcolorbox}

\subsection{
Curating Prompts for Multi-file Editing Data}\label{prompt:multifile}

\begin{tcolorbox}[breakable, colback=blue!5!white, colframe=blue!50!white, title= Data Curation for In-place Edits with Chain Length 2 ]

Based on the above code snippets, complete the following instructions and output according to the format required in Step 3:\
1. Identify interactions across two files:Identify a segment in one file (\#file 1) that is invoked by another file (\#file 2).\
    Exclude import parts in both files and specify only relevant code segments in the corresponding files.\
    2.	Add a new feature:\
	•	Modify the part in \#file 1 that is being invoked by \#file 2.\
	•	Update the corresponding code in \#file 2 to handle the changes in \#file 1.\
	•	If new code snippets are required, append them at the end of each file.\
    3.	Output format:

\noindent\begin{minipage}{\columnwidth}
\begin{lstlisting}[language=json]
{
  "called\_code\_segment": "\#file 1 segment being invoked (excluding `import`)",
  "invoking\_code\_segment": "\#file 2 segment invoking \#file 1 (excluding `import`)",
  "feature\_description": "Description of the new feature",
  "detailed\_feature\_description": "General explanation of the modification approach.",
  "modified\_complete\_code": "Provide the complete code with the required modifications for both files. Include the modified code snippets for \#file 1 and \#file 2. Use comments: \#Modify for modified parts and \#New for newly added parts to indicate whether the change is an addition or modification."
}
\end{lstlisting}
\end{minipage}
\label{Multi-file_Editing_prompt}
\end{tcolorbox}

\begin{tcolorbox}[breakable, colback=blue!5!white, colframe=blue!50!white, title= Data Curation for In-place Edits with Chain Length 3 ]

Based on the above code snippets, complete the following instructions and output according to the format required in Step 3.

1. **Identify interactions across three files:** 

   - Identify a segment in one file (`\#file 1`) that is invoked by another file (`\#file 2`).  
   - Identify how this interaction is further used or invoked in a third file (`\#file 3`).  
   - Check if `\#file 2` contains a segment invoked by `\#file 3`. If so, document this invocation explicitly.  
   - Exclude `import` parts in all cases and specify only relevant code segments in the corresponding files.  

2. **Add a new feature:** 

   - Modify the part in `\#file 1` that is being called by `\#file 2`.  
   - Update `\#file 2` to handle the modified code from `\#file 1` and ensure any related code segments used by `\#file 3` are updated accordingly.  
   - If `\#file 3` directly interacts with updated segments in `\#file 2`, modify the code in `\#file 3` to accommodate the changes.  
   - If new code snippets are required, append them at the end of each file.

3. Output format:

\noindent\begin{minipage}{\columnwidth}
\begin{lstlisting}[language=json]
{
  "called\_code\_segment\_file\_1": "Relevant \#file 1 segment being invoked (excluding `import`)",
  "invoking\_code\_segment\_file\_2": "Relevant \#file 2 segment invoking \#file 1 (excluding `import`)",
  "called\_code\_segment\_file\_2": "Relevant \#file 2 segment being invoked by \#file 3 (excluding `import`)",
  "using\_code\_segment\_file\_3": "Relevant \#file 3 segment using or interacting with \#file 2 or \#file 1 (excluding `import`)",
  "feature\_description": "Description of the new feature",
  "detailed\_feature\_description": "General explanation of the modification approach, including any changes to interactions between \#file 2 and \#file 3.",
  "modified\_complete\_code": "Provide the complete code with the required modifications for all three files. Include the modified code snippets for \#file 1, \#file 2, and \#file 3. Use comments: \#Modify for modified parts and \#New for newly added parts to indicate whether the change is an addition or modification."
}
\end{lstlisting}
\end{minipage}
\label{Multi-file_Editing_prompt}
\end{tcolorbox}

\begin{tcolorbox}[breakable, colback=blue!5!white, colframe=blue!50!white, title= Data Curation for In-place Edits with Chain Length 4 ]

Based on the above code snippets, complete the following instructions and output according to the format required in Step 4.

1. Identify interactions across four files:

	•	Identify a segment in one file (\#file 1) that is invoked by another file (\#file 2).
	•	Identify how this interaction is further used or invoked in a third file (\#file 3).
	•	Check if \#file 2 contains a segment invoked by \#file 3. If so, document this invocation explicitly.
	•	Check if \#file 3 contains a segment invoked by a fourth file (\#file 4). If so, document this invocation explicitly.
	•	Exclude import parts in all cases and specify only relevant code segments in the corresponding files.

2. Add a new feature:

	•	Modify the part in \#file 1 that is being called by \#file 2.
	•	Update \#file 2 to handle the modified code from \#file 1 and ensure any related code segments used by \#file 3 are updated accordingly.
	•	If \#file 3 directly interacts with updated segments in \#file 2, modify the code in \#file 3 to accommodate the changes.
	•	Update \#file 4 if it interacts with or depends on any updated segments in \#file 3.
	•	If new code snippets are required, append them at the end of each file.

3. Output format:

\noindent\begin{minipage}{\columnwidth}
\begin{lstlisting}[language=json]

{
  "called\_code\_segment\_file\_1": "Relevant \#file 1 segment being invoked (excluding `import`)",
  "invoking\_code\_segment\_file\_2": "Relevant \#file 2 segment invoking \#file 1 (excluding `import`)",
  "called\_code\_segment\_file\_2": "Relevant \#file 2 segment being invoked by \#file 3 (excluding `import`)",
  "using\_code\_segment\_file\_3": "Relevant \#file 3 segment using or interacting with \#file 2 or \#file 1 (excluding `import`)",
  "called\_code\_segment\_file\_3": "Relevant \#file 3 segment being invoked by \#file 4 (excluding `import`)",
  "using\_code\_segment\_file\_4": "Relevant \#file 4 segment using or interacting with \#file 3 (excluding `import`)",
  "feature\_description": "Description of the new feature",
  "detailed\_feature\_description": "General explanation of the modification approach, including any changes to interactions between the four files.",
  "modified\_complete\_code": "Provide the complete code with the required modifications for all four files. Include the modified code snippets for \#file 1, \#file 2, \#file 3, and \#file 4. Use comments: \#Modify for modified parts and \#New for newly added parts to indicate whether the change is an addition or modification."
}
\end{lstlisting}
\end{minipage}
\label{Multi-file_Editing_prompt}
\end{tcolorbox}

\begin{tcolorbox}[breakable, colback=blue!5!white, colframe=blue!50!white, title= Data Curation for Expansion Edits with Chain Length 2 ]

Based on the above code snippets, complete the following instructions and output according to the format required in Step 3:

1. **Identify interactions across two files:**

   - Identify a segment in one file (\#file 1) that is invoked by another file (\#file 2).
   - Exclude import parts in both files and specify only relevant code segments in the corresponding files.

2. **Add a new feature:**

   - Modify the part in \#file 1 that is being invoked by \#file 2.
   - Create a new file (`\#file 3`) to implement additional functionality for the new feature. 
   - Update \#file 1 to invoke code from \#file 3 where applicable.
   - Update \#file 2 to invoke code from \#file 3 where applicable.
   - Ensure the changes maintain compatibility across \#file 1 and \#file 2.
   - If new code snippets are required, append them at the end of each file.

3. Output format:

\noindent\begin{minipage}{\columnwidth}
\begin{lstlisting}[language=json]

{
  "called\_code\_segment": "\#file 1 segment being invoked (excluding `import`)",
  "invoking\_code\_segment": "\#file 2 segment invoking \#file 1 (excluding `import`)",
  "new\_file\_code\_segment": "Relevant code snippets added in \#file 3 (excluding `import`)",
  "feature\_description": "Description of the new feature",
  "detailed\_feature\_description": "General explanation of the modification approach, including the purpose and integration of \#file 3.",
  "modified\_complete\_code": "Provide the complete code with the required modifications for \#file 1, \#file 2, and the new file (\#file 3). Use comments: \#Modify for modified parts and \#New for newly added parts to indicate whether the change is an addition or modification."
}
\end{lstlisting}
\end{minipage}
\label{Multi-file_Editing_prompt}
\end{tcolorbox}

\begin{tcolorbox}[breakable, colback=blue!5!white, colframe=blue!50!white, title= Data Curation for Expansion Edits with Chain Length 3 ]
Extended Instruction for Interactions Across Three Files

Based on the above code snippets, complete the following instructions and output according to the format required in Step 3:

	1.	Identify interactions across three files:
	•	Identify a segment in one file (\#file 1) that is invoked by another file (\#file 2).
	•	Identify how this interaction is further used or invoked in a third file (\#file 3).
	•	Check if \#file 2 or \#file 3 contains segments that directly or indirectly depend on \#file 1 or interact with one another. Document these invocations explicitly.
	•	Exclude import parts in all cases and specify only relevant code segments in the corresponding files.

	2.	Add a new feature:
	•	Modify the part in \#file 1 that is being invoked by \#file 2 and/or \#file 3.
	•	Create a new file (\#file 4) to implement additional functionality for the new feature.
	•	Update \#file 1 to invoke code from \#file 4 where applicable.
	•	Update \#file 2 and/or \#file 3 to handle changes in \#file 1 and invoke relevant code from \#file 4.
	•	Ensure compatibility across all interactions involving \#file 1, \#file 2, \#file 3, and the new file (\#file 4).
	•	If new code snippets are required, append them at the end of each file.

3. Output format:

\noindent\begin{minipage}{\columnwidth}
\begin{lstlisting}[language=json]

{
  "called\_code\_segment\_file\_1": "\#file 1 segment being invoked (excluding `import`)",
  "invoking\_code\_segment\_file\_2": "\#file 2 segment invoking \#file 1 (excluding `import`)",
  "invoking\_code\_segment\_file\_3": "\#file 3 segment invoking \#file 1 or \#file 2 (excluding `import`)",
  "new\_file\_code\_segment": "Relevant code snippets added in \#file 4 (excluding `import`)",
  "feature\_description": "Description of the new feature",
  "detailed\_feature\_description": "General explanation of the modification approach, including the purpose and integration of \#file 4.",
  "modified\_complete\_code": "Provide the complete code with the required modifications for \#file 1, \#file 2, \#file 3, and the new file (\#file 4). Use comments: \#Modify for modified parts and \#New for newly added parts to indicate whether the change is an addition or modification."
}
\end{lstlisting}
\end{minipage}

\end{tcolorbox}

\subsection{Prompts for Inference}\label{prompts:inference}

\begin{tcolorbox}[breakable, colback=blue!5!white, colframe=blue!50!white, title=Prompt Template Used for Dependency Recognition]
There are files \#filename. 
Analyze their content and determine the dependency relationship between files. 
Output with the following request. \\
1.Don't give analysis process and using the same file title. \\
2.Simply and only output the dependency relationship list using its file names with the format ['a.py','b.py','c.py'] if b depends on a , c depends on b. \\
3.You must strictly output the response with the format:['a.py','b.py','c.py']
Example output:
["file1.py", "file2.py", "file3.py"]\\
Here's the code snippet:  \#code\_content.
\end{tcolorbox}

\begin{tcolorbox}[breakable, colback=blue!5!white, colframe=blue!50!white, title=Prompt Template Used for Repository Construction]
You are an AI assistant tasked with generating a project structure based on the given repository information. Your task:

1. Analyze the project description, function, and file information provided.

2. Generate a project structure that shows the dependencies between files.

3. Return the structure in the format [[file1, file2, file3], [file4, file5], ...], where each sublist represents a chain of dependencies (file2 calls file1, file3 calls file2, etc.).

4. Provide only the list structure, without any additional explanation.

5. You must strictly follow the output format.

Here's the project information:

Project Description: \#description
Project Function: \#function

Files in the project:
\#files

Based on this information, please generate the project structure showing the dependencies between files. Remember to return only the list structure without any additional explanation.

Example output format:

[["file1.py", "file2.py", "file3.py"], ["file4.py", "file5.py"]]

\end{tcolorbox}

\begin{tcolorbox}[breakable, colback=blue!5!white, colframe=blue!50!white, title=Prompt Template Used for Multi-file Editing]
Based on the above code snippets, complete the following instructions and output according to the format specified in Step 3:\\
1.	Identify the segment in one file that is invoked by another file  (excluding the import parts) and specify the relevant code segment in the called file.
2. Modify the given code to implement the \#function. This requires modifying the part being called. If new code snippets are needed, add them to the end of each respective file.
3. Output format:

\noindent\begin{minipage}{\columnwidth}
\begin{lstlisting}[language=json]

{
   "called\_code\_segment": "\#file 1 segment being invoked (excluding `import`)",
   "invoking\_code\_segment": "\#file 2 segment invoking \#file 1 (excluding `import`)",
   "feature\_description": "Description of the new feature",
   "detailed\_feature\_description": "General explanation of the modification approach",
   "modified\_complete\_code": "Provide the complete code with the required modifications. Output the modified code snippets. Use comments like \#Modify for modified parts and \#New for newly added parts to indicate whether the change is an addition or modification."
}

\end{lstlisting}
\end{minipage}

\end{tcolorbox}

\newpage
\subsection{Prompts for Evaluation}\label{prompts:evaluation}
\begin{tcolorbox}[breakable, colback=blue!5!white, colframe=blue!50!white, title=Prompt Template Used for Evaluating Multi-file Editing]
Gt: \{gt\}
Pred: \{pred\}

Using Gt as the correct answer, compare the content of Pred with Gt and evaluate Pred based on the following aspects. Each aspect contains tailored evaluation criteria to handle the complexities of multi-file interactions and feature integration. The output must follow the JSON format described in Point 6.

Evaluation Aspects

1. Correctness of Function Calls 
Objective: Evaluate the accuracy of all function calls between segments and across files.
	•	Ensure:
	•	Each invoking\_code\_segment correctly calls its corresponding called\_code\_segment as per Gt.
	•	Calls include appropriate parameter matching, order, and context alignment.
	•	Evaluation Criteria:
	•	Does the function signature match, including parameter names, types, and order?
	•	Are correct arguments passed, meeting expectations in feature\_description and detailed\_feature\_description?
	•	Is the pre- or post-logic necessary for context included?
	•	Are cross-file dependencies invoked correctly, as shown in modified\_complete\_code?

Scoring Rules:
	•	5 points: All function calls are completely correct and match Gt, including parameters, order, and logical dependencies.
	•	4 points: Mostly correct with minor parameter or comment issues but no major gaps.
	•	3 points: Partially correct; missing key parameters, logic, or dependencies.
	•	2 points: Significant issues in invocation logic, causing likely runtime errors.
	•	0-1 points: Calls are incorrect, incomplete, or not implemented.

2. Alignment with Feature Requirements
Objective: Check if the code in Pred aligns with the intended feature and modification goals.
	•	Ensure:
	•	Every call reflects requirements in feature\_description and detailed\_feature\_description.
	•	The new or modified logic directly implements the required functionality.
	•	Evaluation Criteria:
	•	Does the logic adhere to the functional goals described?
	•	Does it integrate with multi-file dependencies correctly (if applicable)?
	•	Are the new components in new\_file\_code\_segment aligned with expectations?

Scoring Rules:
	•	5 points: Perfectly aligned with feature requirements; implementation is logically complete.
	•	4 points: Correctly aligned but with potential optimizations or minor improvements.
	•	3 points: Partially fulfills requirements with clear gaps in alignment.
	•	2 points: Loosely aligned with significant logic missing.
	•	0-1 points: Not aligned or entirely unrelated to the described requirements.

3. Accuracy of Functionality Implementation
Objective: Verify the correctness of the implementation, focusing on functional outcomes.
	•	Evaluation Criteria:
	•	Does the functionality fully satisfy the requirements in feature\_description?
	•	Are components correctly loaded, initialized, or referenced?
	•	Are all dependencies resolved for seamless multi-file integration?

Scoring Rules:
	•	5 points: Fully accurate implementation without functional defects.
	•	4 points: Mostly accurate with minor issues or deviations.
	•	3 points: Partially correct but lacking essential steps or logic.
	•	2 points: Basic framework present but largely incomplete.
	•	0-1 points: Non-functional due to missing or incorrect logic.

4. Completeness of Implementation
Objective: Ensure that all functional components, including new and modified ones, are fully implemented.
	•	Evaluation Criteria:
	•	Are all required segments across files defined and updated per Gt?
	•	Does the implementation cover all subparts described in detailed\_feature\_description?
	•	Are all new dependencies (\#New segments) and modifications (\#Modify segments) accounted for?

Scoring Rules:
	•	5 points: Complete implementation with no omissions.
	•	4 points: Nearly complete, with only minor omissions.
	•	3 points: Significant missing functionality, but partially meets requirements.
	•	2 points: Too many missing components, achieving minimal functionality.
	•	0-1 points: Nearly all components are missing or incorrect.

5. Code Quality
Objective: Assess the overall quality, maintainability, and readability of the code.
	•	Evaluation Criteria:
	•	Readability: Clear naming, concise comments, and consistent style.
	•	Maintainability: Modular structure, minimal duplication, and extensibility.
	•	Efficiency: Appropriate algorithms, data structures, and resource use.

Scoring Rules:
	•	5 points: Excellent quality with clean, efficient, and maintainable code.
	•	4 points: Good quality, but minor readability or efficiency issues.
	•	3 points: Average quality; readable but not optimized or modular.
	•	2 points: Poor quality; lacks structure or suffers from inefficiencies.
	•	0-1 points: Unreadable, unstructured, or inefficient code.

6. JSON Output Format

\noindent\begin{minipage}{\columnwidth}
\begin{lstlisting}[language=json]

{
  "correctness_score": 5,
  "purpose_alignment_score": 4,
  "functionality_accuracy_score": 5,
  "functionality_completeness_score": 4,
  "code_quality_score": 5
}

\end{lstlisting}
\end{minipage}
\end{tcolorbox}

\newpage
\section{Task Case}

\begin{tcolorbox}[breakable, colback=blue!5!white, colframe=blue!50!white, title=Instance of Dependency Recognition]
"files":["'bilireq/bilireq/login/\_init\_.py'", "'bilireq/bilireq/typing.py'", "'bilireq/test/test\_login.py'"],\\
"content":
\begin{lstlisting}[language=Python]
import asyncio
from base64 import b64encode
from io import BytesIO
from typing import Optional, Union

from qrcode.image.pure import PyPNGImage
from qrcode.main import QRCode
from _typing import T_Auth
from auth import Auth, WebAuth
from ..exceptions import ResponseCodeError
from ..utils import get, post
from .pwd_login import pwd_login as _pwd_login
from .qrcode_login import get_qrcode_login_info, get_qrcode_login_result
from .sms_login import send_sms
from .sms_login import sms_login as _sms_login
from .web_qrcode_login import get_web_qrcode_login_info, get_web_qrcode_login_url

BASE_URL = "https://passport.bilibili.com/api/v2/oauth2/"

async def refresh_token(auth: T_Auth = None, *, reqtype="app", **kwargs):
    url = f"{BASE_URL}refresh_token"
    return await post(url, auth=auth, reqtype=reqtype, **kwargs)

async def get_token_info(auth: T_Auth = None, *, reqtype="app", **kwargs):
    url = f"{BASE_URL}info"
    return await get(url, auth=auth, reqtype=reqtype, **kwargs)

class Login:
    auth_code: str
    qrcode_url: str
    tel: int
    cid: int
    captcha_key: str

    async def get_web_qrcode_url(self) -> str:
        r = await get_web_qrcode_login_url()
        self.auth_code = r["qrcode_key"]
        self.qrcode_url = r["url"]
        return self.qrcode_url

    async def get_qrcode_url(self) -> str:
        r = await get_qrcode_login_info()
        self.auth_code = r["auth_code"]
        self.qrcode_url = r["url"]
        return self.qrcode_url

    async def get_qrcode(self, url: Optional[str] = None, 
                         print_qr=False, base64_encode=False, 
        url = url or (await self.get_qrcode_url() 
                      if login_type == "app" 
                      else await self.get_web_qrcode_url())

        qr = QRCode()
        qr.add_data(url)

        if print_qr:
            qr.print_tty()
            return None

        img = qr.make_image(image_factory=PyPNGImage)
        buf = BytesIO()
        img.save(buf)

        if not base64_encode:
            return buf.getvalue()
        return b64encode(buf.getvalue()).decode()

    async def web_qrcode_login(self, auth_code=None, retry=-1, interval=1):
        auth_code = auth_code or self.auth_code
        # Implementation continues...
\end{lstlisting}
"gt":["'bilireq/bilireq/\_typing.py'",
"'bilireq/bilireq/login/\_init\_.py'",
"'bilireq/test/test\_login.py'"
]

\end{tcolorbox}

\begin{tcolorbox}[breakable, colback=blue!5!white, colframe=blue!50!white, title=Instance of Repository Construction]
"repo": "got-your-back", \\
"description": "GYB is a command-line tool for backing up Gmail messages to a local computer.",\\
"function": "Utilizes Gmail's API over HTTPS to securely download and store emails locally, offering installation options for Linux, MacOS, and Windows.",\\
"files": [
    \{
    "file": "got-your-back/fmbox.py",
    "function": "This library provides functionality to read and manipulate mbox files sequentially, allowing extraction, modification, and removal of email headers, as well as iterating through messages in an mbox file."
    \},
    {
    "file": "got-your-back/gyb.py",
    "function": "This script is a command-line tool for backing up and restoring Gmail messages. It supports various actions such as backup, restore, count, purge, and label management, and integrates with Google APIs for Gmail and Google Workspace services. The tool uses OAuth 2.0 for authentication and supports both user accounts and service accounts."
    },
    {
    "file": "got-your-back/labellang.py",
    "function": "Unable to read file content."
    }
],\\
"gt": "[['got-your-back/labellang.py', 'got-your-back/gyb.py'], ['got-your-back/fmbox.py', 'got-your-back/gyb.py']]"

\end{tcolorbox}

\begin{tcolorbox}[breakable, colback=blue!5!white, colframe=blue!50!white, title=Instance of Repository Construction, label=note_taxonomy_single]

"repo": "ip\_basic",\\

        "content": 
\begin{lstlisting}[language=python]
'import ip_basic/ip_basic/vis_utils.py'
import cv2

def cv2_show_image(window_name, image, size_wh=None, location_xy=None):
    """Helper function for specifying window size and location when displaying images with cv2.

    Args:
        window_name: str window name
        image: ndarray image to display
        size_wh: window size (w, h)
        location_xy: window location (x, y)
    """

    if size_wh is not None:
        cv2.namedWindow(window_name, cv2.WINDOW_KEEPRATIO | cv2.WINDOW_GUI_NORMAL)
        cv2.resizeWindow(window_name, *size_wh)
    else:
        cv2.namedWindow(window_name, cv2.WINDOW_AUTOSIZE)

    if location_xy is not None:
        cv2.moveWindow(window_name, *location_xy)

    cv2.imshow(window_name, image)

'import ip_basic/demos/depth_completion.py'
import glob
import os
import sys
import time
import cv2
import numpy as np
import png
from ip_basic import depth_map_utils
from ip_basic import vis_utils

def main():
    """Depth maps are saved to the 'outputs' folder."""
    ##############################
    # Options
    ##############################
    # Validation set
    input_depth_dir = os.path.expanduser('~/Kitti/depth/depth_selection/val_selection_cropped/velodyne_raw')
    data_split = 'val'

    # Test set
    # input_depth_dir = os.path.expanduser('~/Kitti/depth/depth_selection/test_depth_completion_anonymous/velodyne_raw')
    # data_split = 'test'

    # Fast fill with Gaussian blur @90Hz (paper result)
    fill_type = 'fast'
    extrapolate = True
    blur_type = 'gaussian'

    # Fast Fill with bilateral blur, no extrapolation @87Hz (recommended)
    # fill_type = 'fast'
    # extrapolate = False
    # blur_type = 'bilateral'

    # Multi-scale dilations with extra noise removal, no extrapolation @ 30Hz
    # fill_type = 'multiscale'
    # extrapolate = False
    # blur_type = 'bilateral'

    # Save output to disk or show process
    save_output = True

    ##############################
    # Processing
    ##############################
    if save_output:
        # Save to Disk
        show_process = False
        save_depth_maps = True
    else:
        if fill_type == 'fast':
            raise ValueError('"fast" fill does not support show_process')

        # Show Process
        show_process = True
        save_depth_maps = False

    # Create output folder
    this_file_path = os.path.dirname(os.path.realpath(__file__))
    outputs_dir = this_file_path + '/outputs'
    os.makedirs(outputs_dir, exist_ok=True)

    output_folder_prefix = 'depth_' + data_split
    output_list = sorted(os.listdir(outputs_dir))
    if len(output_list) > 0:
        split_folders = [folder for folder in output_list if folder.startswith(output_folder_prefix)]
        if len(split_folders) > 0:
            last_output_folder = split_folders[-1]
            last_output_index = int(last_output_folder.split('_')[-1])
        else:
            last_output_index = -1
    else:
        last_output_index = -1
    output_depth_dir = outputs_dir + '/{}_{:03d}'.format(output_folder_prefix, last_output_index + 1)

    if save_output:
        if not os.path.exists(output_depth_dir):
            os.makedirs(output_depth_dir)
        else:
            raise FileExistsError('Already exists!')
        print('Output dir:', output_depth_dir)

    # Get images in sorted order
    images_to_use = sorted(glob.glob(input_depth_dir + '/*'))

    # Rolling average array of times for time estimation
    avg_time_arr_length = 10
    last_fill_times = np.repeat([1.0], avg_time_arr_length)
    last_total_times = np.repeat([1.0], avg_time_arr_length)

    num_images = len(images_to_use)
    for i in range(num_images):

        depth_image_path = images_to_use[i]

        # Calculate average time with last n fill times
        avg_fill_time = np.mean(last_fill_times)
        avg_total_time = np.mean(last_total_times)

        # Show progress
        sys.stdout.write('\rProcessing {} / {}, '
                         'Avg Fill Time: {:.5f}s, '
                         'Avg Total Time: {:.5f}s, '
                         'Est Time Remaining: {:.3f}s'.format(
                             i, num_images - 1, avg_fill_time, avg_total_time,
                             avg_total_time * (num_images - i)))
        sys.stdout.flush()

        # Start timing
        start_total_time = time.time()

        # Load depth projections from uint16 image
        depth_image = cv2.imread(depth_image_path, cv2.IMREAD_ANYDEPTH)
        projected_depths = np.float32(depth_image / 256.0)

        # Fill in
        start_fill_time = time.time()
        if fill_type == 'fast':
            final_depths = depth_map_utils.fill_in_fast(
                projected_depths, extrapolate=extrapolate, blur_type=blur_type)
        elif fill_type == 'multiscale':
            final_depths, process_dict = depth_map_utils.fill_in_multiscale(
                projected_depths, extrapolate=extrapolate, blur_type=blur_type,
                show_process=show_process)
        else:
            raise ValueError('Invalid fill_type {}'.format(fill_type))
        end_fill_time = time.time()

        # Display images from process_dict
        if fill_type == 'multiscale' and show_process:
            img_size = (570, 165)

            x_start = 80
            y_start = 50
            x_offset = img_size[0]
            y_offset = img_size[1]
            x_padding = 0
            y_padding = 28

            img_x = x_start
            img_y = y_start
            max_x = 1900

            row_idx = 0
            for key, value in process_dict.items():

                image_jet = cv2.applyColorMap(
                    np.uint8(value / np.amax(value) * 255),
                    cv2.COLORMAP_JET)
                vis_utils.cv2_show_image(
                    key, image_jet,
                    img_size, (img_x, img_y))

                img_x += x_offset + x_padding
                if (img_x + x_offset + x_padding) > max_x:
                    img_x = x_start
                    row_idx += 1
                img_y = y_start + row_idx * (y_offset + y_padding)

                # Save process images
                cv2.imwrite('process/' + key + '.png', image_jet)

            cv2.waitKey()

        # Save depth images to disk
        if save_depth_maps:
            depth_image_file_name = os.path.split(depth_image_path)[1]

            # Save depth map to a uint16 png (same format as disparity maps)
            file_path = output_depth_dir + '/' + depth_image_file_name
            with open(file_path, 'wb') as f:
                depth_image = (final_depths * 256).astype(np.uint16)

                # pypng is used because cv2 cannot save uint16 format images
                writer = png.Writer(width=depth_image.shape[1],
                                    height=depth_image.shape[0],
                                    bitdepth=16,
                                    greyscale=True)
                writer.write(f, depth_image)

        end_total_time = time.time()

        # Update fill times
        last_fill_times = np.roll(last_fill_times, -1)
        last_fill_times[-1] = end_fill_time - start_fill_time

        # Update total times
        last_total_times = np.roll(last_total_times, -1)
        last_total_times[-1] = end_total_time - start_total_time


if __name__ == "__main__":
    main()
\end{lstlisting}

"called\_code\_segment": 
    
\begin{lstlisting}[language=python]
def cv2_show_image(window_name, image, size_wh=None, location_xy=None):
    """Helper function for specifying window size and location when
    displaying images with cv2.

    Args:
        window_name: str window name
        image: ndarray image to display
        size_wh: window size (w, h)
        location_xy: window location (x, y)
    """
    if size_wh is not None:
        cv2.namedWindow(window_name,
                        cv2.WINDOW_KEEPRATIO | cv2.WINDOW_GUI_NORMAL)
        cv2.resizeWindow(window_name, *size_wh)
    else:
        cv2.namedWindow(window_name, cv2.WINDOW_AUTOSIZE)

    if location_xy is not None:
        cv2.moveWindow(window_name, *location_xy)

    cv2.imshow(window_name, image)
\end{lstlisting}
    
        "feature\_description": "Add an option to save the displayed images to disk.",\\
        "detailed\_feature\_description": "Modify the `cv2\_show\_image` function in \#file 1 to accept an additional parameter `save\_path` which, if provided, will save the displayed image to the specified path. Update the invocation in \#file 2 to include this new parameter and save the images to a specified directory.",\\
        "modified\_complete\_code": 
        
        \begin{lstlisting}[language=python]
# file 1
import cv2

def cv2_show_image(window_name, image, size_wh=None, location_xy=None, save_path=None):
    """
    Helper function for specifying window size and location when displaying images with cv2.

    Args:
        window_name: str, window name
        image: ndarray, image to display
        size_wh: tuple, window size (w, h)
        location_xy: tuple, window location (x, y)
        save_path: str, path to save the image
    """
    if size_wh is not None:
        cv2.namedWindow(window_name, cv2.WINDOW_KEEPRATIO | cv2.WINDOW_GUI_NORMAL)
        cv2.resizeWindow(window_name, *size_wh)
    else:
        cv2.namedWindow(window_name, cv2.WINDOW_AUTOSIZE)

    if location_xy is not None:
        cv2.moveWindow(window_name, *location_xy)

    cv2.imshow(window_name, image)

    if save_path is not None:
        cv2.imwrite(save_path, image)
        

# file 2
if fill_type == 'multiscale' and show_process:
    img_size = (570, 165)

    x_start = 80
    y_start = 50
    x_offset = img_size[0]
    y_offset = img_size[1]
    x_padding = 0
    y_padding = 28

    img_x = x_start
    img_y = y_start
    max_x = 1900

    row_idx = 0
    for key, value in process_dict.items():
        image_jet = cv2.applyColorMap(np.uint8(value / np.amax(value) * 255), cv2.COLORMAP_JET)
        save_path = 'process/' + key + '.png'
        vis_utils.cv2_show_image(key, image_jet, img_size, (img_x, img_y), save_path=save_path)

        img_x += x_offset + x_padding
        if (img_x + x_offset + x_padding) > max_x:
            img_x = x_start
            row_idx += 1
        img_y = y_start + row_idx * (y_offset + y_padding)

    cv2.waitKey()
\end{lstlisting}
    
\end{tcolorbox}
\section{Crowdsourcing}
\label{app:crowdsourcing}

In conducting our study, we identified several potential risks to participants. Firstly, there is a risk to privacy and confidentiality, as participants are required to share personal information. To mitigate this, all data will be anonymized and stored securely, with access restricted to authorized personnel only. Secondly, there may be psychological risks, such as discomfort or stress during the tasks. To address this, we have included detailed instructions and debriefing sessions to ensure participants feel supported throughout the process. Additionally, participants have the right to withdraw from the study at any time without penalty. Lastly, while there are no significant physical risks associated with our procedures, we will monitor participants for any signs of distress and provide appropriate support. We pay each participant an hourly rate of \$10. The primary participants we recruit are college students majored in software engineering with master degree (Age ranging 23-28).

\end{document}